\documentclass[12pt]{article}
\usepackage{amssymb}

\def\hybrid{\topmargin 0pt      \oddsidemargin 0pt
        \headheight 0pt \headsep 0pt
        \voffset=-0.5cm
        \hoffset=-0.25in
        \textwidth 6.75in
        \textheight 9.5in       
        \marginparwidth 0.0in
        \parskip 5pt plus 1pt   \jot = 1.5ex}
\catcode`\@=11
\def\marginnote#1{}

\newcount\hour
\newcount\minute
\newtoks\amorpm
\hour=\time\divide\hour by60 \minute=\time{\multiply\hour by60 \global\advance\minute by-\hour}
\edef\standardtime{{\ifnum\hour<12 \global\amorpm={am}%
        \else\global\amorpm={pm}\advance\hour by-12 \fi
        \ifnum\hour=0 \hour=12 \fi
        \number\hour:\ifnum\minute<10 0\fi\number\minute\the\amorpm}}
\edef\militarytime{\number\hour:\ifnum\minute<10 0\fi\number\minute}

\def\draftlabel#1{{\@bsphack\if@filesw {\let\thepage\relax
   \xdef\@gtempa{\write\@auxout{\string
      \newlabel{#1}{{\@currentlabel}{\thepage}}}}}\@gtempa
   \if@nobreak \ifvmode\nobreak\fi\fi\fi\@esphack}
        \gdef\@eqnlabel{#1}}
\def\@eqnlabel{}
\def\@vacuum{}
\def\draftmarginnote#1{\marginpar{\raggedright\scriptsize\tt#1}}
\def\draftlabel#1{{\@bsphack\if@filesw {\let\thepage\relax
   \xdef\@gtempa{\write\@auxout{\string
      \newlabel{#1}{{\@currentlabel}{\thepage}}}}}\@gtempa
   \if@nobreak \ifvmode\nobreak\fi\fi\fi\@esphack}
        \gdef\@eqnlabel{#1}}
\def\@eqnlabel{}
\def\@vacuum{}
\def\draftmarginnote#1{\marginpar{\raggedright\scriptsize\tt#1}}

\def\draft{\oddsidemargin -.5truein
        \def\@oddfoot{\sl preliminary draft \hfil
        \rm\thepage\hfil\sl\today\quad\militarytime}
        \let\@evenfoot\@oddfoot \overfullrule 3pt
        \let\label=\draftlabel
        \let\marginnote=\draftmarginnote
   \def\@eqnnum{(\theequation)\rlap{\kern\marginparsep\tt\@eqnlabel}%
\global\let\@eqnlabel\@vacuum}  }

\def\numberbysection{\@addtoreset{equation}{section}
        \def\theequation{\thesection.\arabic{equation}}}

\def\underline#1{\relax\ifmmode\@@underline#1\else
        $\@@underline{\hbox{#1}}$\relax\fi}

\def\titlepage{\@restonecolfalse\if@twocolumn\@restonecoltrue\onecolumn
     \else \newpage \fi \thispagestyle{empty}\c@page\z@
        \def\thefootnote{\fnsymbol{footnote}} }

\def\endtitlepage{\if@restonecol\twocolumn \else  \fi
        \def\thefootnote{\arabic{footnote}}
        \setcounter{footnote}{0}}  

\numberbysection \hybrid

\newcounter{mo}

\newcommand{\tr}{{\rm tr}}
\newcommand{\ti}[1]{\tilde{#1}}

\newcommand{\vf}{\varphi}
\newcommand{\al}{\alpha}
\newcommand{\be}{\beta}

\newcommand{\om}{\omega}
\newcommand{\vth}{\vartheta}

\newcommand{\bfe}{{\bf{e}}}

\newtheorem{predl}{Proposition}[section]

\newtheorem{rem}{Remark}
\newtheorem{defi}{Definition}

\def\beq{\begin{equation}}
\def\eq{\end{equation}}
\def\p{\partial}

\newcommand{\mat}[4]{\left(\begin{array}{cc}{#1}&{#2}\\ \ \\{#3}&{#4}
\end{array}\right)}

\def\res{\mathop{\hbox{Res}}\limits}

\begin{document}

\setcounter{page}{1}

\date{}
\date{}
\vspace{50mm}

\begin{flushright}
 ITEP-TH-27/14\\
\end{flushright}
\vspace{0mm}

\begin{center}
\vspace{0mm}
%
{\LARGE{Planck Constant as Spectral Parameter}}
 \\ \vspace{4mm}
  {\LARGE{in Integrable Systems and KZB Equations}}
\\
\vspace{12mm} {\large {A. Levin}\,$^{\flat\,\sharp}$ \ \ \ \ {M.
Olshanetsky}\,$^{\sharp\,\natural}$
 \ \ \ {A. Zotov}\,$^{\diamondsuit\, \sharp\, \natural}$ }\\
 \vspace{8mm}

 \vspace{2mm} $^\flat$ -- {\small{\sf 
 NRU HSE, Department of Mathematics,
 Myasnitskaya str. 20,  Moscow,  101000,  Russia}}\\
 \vspace{2mm} $^\sharp$ -- {\small{\sf 
 ITEP, B. Cheremushkinskaya str. 25,  Moscow, 117218, Russia}}\\
 \vspace{2mm} $^\natural$ -- {\small{\sf MIPT, Inststitutskii per.  9, Dolgoprudny,
 Moscow region, 141700, Russia}}\\
\vspace{2mm} $^\diamondsuit$ -- {\small{\sf Steklov Mathematical
Institute  RAS, Gubkina str. 8, Moscow, 119991,  Russia}}\\
\end{center}

\begin{center}\footnotesize{{\rm E-mails:}{\rm\ \
 alevin@hse.ru,\  olshanet@itep.ru,\  zotov@mi.ras.ru}}\end{center}

 \begin{abstract}
We construct special  rational ${\rm gl}_N$
Knizhnik-Zamolodchikov-Bernard (KZB) equations with $\tilde N$
punctures by deformation of the corresponding quantum ${\rm gl}_N$
rational $R$-matrix. They have two parameters. The limit of the
first one brings the model to the ordinary rational KZ equation.
Another one is $\tau$. At the level of classical mechanics the
deformation parameter $\tau$ allows to extend the previously
obtained  modified Gaudin models to the modified Schlesinger
systems. Next, we notice that the identities underlying  generic
(elliptic) KZB equations follow from some additional relations for
the properly normalized $R$-matrices. The relations are
noncommutative analogues of identities for (scalar) elliptic
functions. The simplest one is the unitarity condition. The
quadratic (in $R$ matrices) relations are generated by
noncommutative Fay identities. In particular, one can derive the
quantum Yang-Baxter equations from the Fay identities. The cubic
relations provide identities for the KZB equations as well as
quadratic relations for the classical $r$-matrices which can be
halves of the classical Yang-Baxter equation. At last we discuss the
$R$-matrix valued linear problems which provide ${\rm gl}_{\ti N}$
Calogero-Moser (CM) models and Painlev\'e equations via the above
mentioned identities. The role of the spectral parameter plays the
Planck constant of the quantum $R$-matrix. When the quantum ${\rm
gl}_N$ $R$-matrix is scalar ($N=1$) the linear problem reproduces
the Krichever's ansatz for the Lax matrices with spectral parameter
for the ${\rm gl}_{\ti N}$ CM models.  The linear problems for the
quantum CM models
 generalize the KZ equations in the same way as the Lax
pairs with spectral parameter generalize those without it.
 \end{abstract}

\newpage

{\small{

\tableofcontents

}}


\section{Introduction}
\setcounter{equation}{0}

Let $V$ be a finite-dimensional module of the group ${\rm GL}_N$.
The quantum $R$-matrix is an operator $R\,:\,V\otimes V\to V\otimes
V$ satisfying the quantum Yang-Baxter equation \cite{Sklyanin0}:
 \beq\label{q102}
 \begin{array}{c}
  \displaystyle{
 R^\hbar_{12}(z-w)\,R^\hbar_{13}(z)\,R^\hbar_{23}(w)=R^\hbar_{23}(w)\,
 R^\hbar_{13}(z)\, R^\hbar_{12}(z-w)\,,
 }
 \end{array}
 \eq
 where $z,w$ - spectral parameters.
 We consider a special class of non-dynamical $R$-matrices which
includes Belavin's elliptic ${\rm gl}_N$ $R$-matrix and its
(nontrivial) degenerations, i.e. $z$ is a local coordinate on the
(degenerated) elliptic curve. Let us fix the normalization of
$R^\hbar$ in the way that the unitarity condition takes the form
  \beq\label{q1021}
 \begin{array}{c}
  \displaystyle{
 R^\hbar_{12}(z)R^\hbar_{21}(z)=1\otimes 1
 \Phi^\hbar(z)\Phi^\hbar(-z)\,,
 }
 \end{array}
 \eq
 where $\Phi^\hbar(z)$ is the function defined in the elliptic
 case\footnote{In the rational case we use $\Phi^\hbar(z)=z^{-1}+\hbar^{-1}$. The
trigonometric case will be considered separately.}
 as
     \beq\label{iq1498}
 \begin{array}{c}
  \displaystyle{
\Phi^\hbar(z)=N\phi(N\hbar,z)\,,\ \ \ \phi(z,u)=\frac
{\vth'(0)\vth(u+z)} {\vth(z)\vth(u)}\,,
 }
 \end{array}
 \eq
where $\vth(z)=\theta_{11}(z|\tau)$ is the odd Riemann
theta-function, $\tau$ -- elliptic moduli.

We demonstrate here that starting with the $R$-matrix one can
construct different
 families of classical and quantum integrable system. These constructions are based
 on two special features of the $R$-matrices. The first one is the quasi-classical
 expansion. With the normalization (\ref{q1021})-(\ref{iq1498}) it acquires the form:
 \beq\label{q104}
 \begin{array}{c}
  \displaystyle{
 R^\hbar_{12}(z)=\frac{1}{\hbar}\,1\otimes 1+r_{12}(z)+\hbar\,\,
 m_{12}(z)+O(\hbar^2)\,,
 }
 \end{array}
 \eq
where $r_{12}(z)$ is the classical $r$-matrix.
 It leads to integrable Euler-Arnold
 ${\rm gl}_N$
 tops\footnote{The integrable tops were previously proved to be related (equivalent) to the (spin)
 Calogero-Ruijsenaars
 models
 by the symplectic Hecke transformations. See. e.g. \cite{LOZ,LOZ5,LOZ7}} and Gaudin systems.

 The second is the property of Painlev\'e-Calogero
 correspondence, which is equivalent to the heat equation:
  \beq\label{q030}
   \begin{array}{|c|}
  \hline\\
 \displaystyle{
\p_\tau R_{12}^\hbar(z)=\p_z\p_\hbar R_{12}^\hbar(z)
  }
   \\ \ \\
 \hline
 \end{array}
 \eq
 The latter leads to the monodromy preserving equations (non-autonomous tops, Schlesinger
 systems) and the KZB systems.

 At last, the main tool is the set of
 identities for the quantum $R$-matrices which we introduce below.
$R$-matrix is an operator acting on the tensor product of vector
spaces $V$. Consider a set of points $z_1,...,z_{\ti N}$ (on the
curve where $z$ is a local coordinate). Let
   \beq\label{q042}
 \begin{array}{c}
  \displaystyle{
R^\hbar_{ab}=R^\hbar(z_a-z_b)\,,
 }
 \end{array}
 \eq
be the $R$-matrix acting on the $a$-th and $b$-th components of
$V^{\otimes \ti N}$. In our case $R$-matrices satisfy the following
property:
   \beq\label{q1492}
 \begin{array}{c}
  \displaystyle{
R^\hbar_{ab}(z_a-z_b)=-R^{-\hbar}_{ba}(z_b-z_a)\,,
 }
 \end{array}
 \eq
i.e. the terms of the expansion (\ref{q104}) are of definite parity:
   \beq\label{q043}
 \begin{array}{c}
  \displaystyle{
r_{ab}=-r_{ba}\,,\ \ \ m_{ab}=m_{ba}\,.
 }
 \end{array}
 \eq
We show that the $R$-matrices satisfy a set of relations similar to
identities for function $\phi(z,u)$ (\ref{iq1498}). In particular,
$\phi(z,u)$ satisfies the Fay identity
  \beq\label{iq701}
 \begin{array}{c}
  \displaystyle{
 \phi(x,z_{ab})\phi(y,z_{bc})=\phi(x-y,z_{ab})\phi(y,z_{ac})+\phi(y-x,z_{bc})\phi(x,z_{ac})\,,
 }
 \end{array}
 \eq
where $z_{ab}=z_a-z_b$. We notice that the following analogue of the
Fay identity holds:
 \beq\label{iq702}
   \begin{array}{|c|}
  \hline\\
 \displaystyle{
R^\hbar_{ab}
 R^{\hbar'}_{bc}=R^{\hbar'}_{ac}R_{ab}^{\hbar-\hbar'}+R^{\hbar'-\hbar}_{bc}R^\hbar_{ac}
  }
   \\ \ \\
 \hline
 \end{array}
 \eq
It will be shown that one can derive the quantum Yang-Baxter
equation  (\ref{q102}) from (\ref{iq702}).

While the quantum $R$-matrix is similar to $\phi(\hbar,z)$ the
classical $r$-matrix is the analogue of function
$E_1(z)=\p_z\log\vth(z)$. For example, the following relation holds:
 \beq\label{iq409}
 \begin{array}{c}
  \displaystyle{
(r_{ab}+r_{bc}+r_{ca})^2=1_a\otimes 1_b\otimes 1_c\,
N^2(\wp(z_a-z_b)+\wp(z_b-z_c)+\wp(z_c-z_a))\,,
 }
 \end{array}
 \eq
where $\wp(z)$ is the Weierstrass  $\wp$-function with moduli
$\tau$. It is the analogue of the identity
 \beq\label{iq410}
 \begin{array}{c}
  \displaystyle{
(E_1(z_a-z_b)+E_1(z_b-z_c)+E_1(z_c-z_a))^2=\wp(z_a-z_b)+\wp(z_b-z_c)+\wp(z_c-z_a)\,.
 }
 \end{array}
 \eq
Together with (\ref{iq409}) the classical Yang-Baxter equation
 \beq\label{iq406}
 \begin{array}{c}
  \displaystyle{
[r_{ab},r_{ac}]+[r_{ac},r_{bc}]+[r_{ab},r_{bc}]=0
 }
 \end{array}
 \eq
leads to the following relations:
 \beq\label{iq407}
   \begin{array}{|c|}
  \hline\\
 \displaystyle{
r_{ab}\,r_{ac}-r_{bc}\,r_{ab}+r_{ac}\,r_{bc}=m_{ab}+m_{bc}+m_{ac}\,.
  }
   \\ \ \\
 \hline
 \end{array}
 \eq
Difference of (\ref{iq407}) written for indices $a,b,c$ and $a,c,b$
gives (\ref{iq406}).

Let us remark that the class of $R$-matrices we discuss here
includes Baxter-Belavin's one \cite{Baxter,Belavin} as the most
general. Its trigonometric analogue was found in
\cite{Cherednik1,Zabrodin1} (we are going to consider it in separate
publications). At last the rational case is known from
\cite{Cherednik1,Smirnov1,LOZ7}. In the simplest cases one gets the
ordinary XXZ and XXX Yang's $R$-matrices. In the rational case the
Yang's $R$-matrix \cite{Yang} (with normalization (\ref{q1021})) is
of the form:
 \beq\label{iq419}
 \begin{array}{c}
  \displaystyle{
R^{\hbar,\hbox{\tiny{Yang}}}_{ab}=\frac{1_a\otimes
1_b}{\hbar}+\frac{P_{ab}}{z_a-z_b}\,,
 }
 \end{array}
 \eq
where $P_{ab}$ is the permutation operator. We deal with non-trivial
deformations of (\ref{iq419}). In particular, they allow us to
define not only KZ but also KZB equations. At the same time the rest
of our construction works for ordinary XXX (and XXZ) $R$-matrices as
well\footnote{It is interesting if similar construction works for
Toda-like models which can be obtained from the elliptic systems by
nontrivial (Inozemtsev) degenerations.}.

{\bf The purpose of the paper} is twofold. First, we construct the
rational analogue of the (elliptic) KZB equations. For this purpose
we find $\tau$ deformation of the quantum $R$-matrix suggested in
\cite{LOZ7}. Second, we show that integrable systems of
Calogero-Moser type admit higher rank Lax representations which
generalize the Krichever's one \cite{Krich1} in the same way as
(\ref{iq702}) generalize (\ref{iq701}). The standard (non-diagonal)
matrix elements $\phi(\lambda,z_a-z_b)$ are replaced by the quantum
$R$-matrices $R^\lambda_{ab}$, i.e. the spectral parameter is given
by the Planck constant entering $R$-matrix.
Our constructions are independent of specific form of the
 $R$-matrix, but based only on the set of identities (such as (\ref{iq702}), (\ref{iq407}), (\ref{q030})) which can be
 verified separately.

\vskip2mm

{\bf 1. Rational KZB equations}

\vskip2mm

 Besides the standard trigonometric and rational versions of the
elliptic $R$-matrix there are more sophisticated degenerations.
 In this paper we consider one of them \cite{LOZ7} and show that it leads to
  some modifications of the standard Gaudin
 and Schlesinger systems
 and the KZ (KZB) equations.

The Belavin's $R$-matrix depends on the moduli of the elliptic curve
$\tau$. We notice that it satisfies the heat equation (\ref{q030})
and treat this equation as Painev\'e-Calogero property. In \cite{LO}
it was formulated in the following way: the Lax pair of the CM model
satisfies also the monodromy preserving equations and describe the
(higher rank) Painlev\'e equations. We refer to (\ref{q030}) as the
heat equation because this equation for the function $\phi(\hbar,z)$
follows from the heat equation for $\vth$-function
$2\p_\tau\vth(z|\tau)=\p_z^2\vth(z|\tau)$.

 The natural (noncommutative)
analogue of $\vth$-function is the modification of bundle
$\Xi(z,\tau)$. In the elliptic case it was found in
\cite{Hasegawa12} in the context of the IRF-Vertex transformation,
and then described in \cite{LOZ} (see also \cite{LOZ5,LOZ7}) as an
example of the Symplectic Hecke Correspondence for integrable
systems. Its rational analogue was suggested in \cite{AASZ} and was
know to be free of $\tau$ dependence. Here we explain how to
introduce the $\tau$-dependence.
We construct the $\tau$ deformation of the rational $R$-matrix based
on the heat equation
 \beq\label{q032}
 \begin{array}{c}
  \displaystyle{
2\p_\tau\Xi=\p^2_z\Xi\,.
 }
 \end{array}
 \eq
The solution provides possibility for construction of the rational
analogue of the KZB equations
 \beq\label{iq140}
 \left\{
 \begin{array}{l}
  \displaystyle{
\hat\nabla_a \psi=0\,,\ \ \ \nabla_a=\p_{z_a}+\sum\limits_{c\neq a}
{\mathfrak r}^\tau_{ac}(z_a-z_c)\,,
 }
 \\ \ \\
   \displaystyle{
\hat\nabla_\tau \psi=0\,,\ \ \
\nabla_\tau=\p_\tau+\frac{1}{2}\sum\limits_{b,c} {\mathfrak
m}^\tau_{bc}(z_b-z_c)\,,
 }
 \end{array}
 \right.
 \eq
where $r$ and $m$ are the terms of the expansion (\ref{q104}) and
$\tau$ indicates the $\tau$-deformation. The system of KZ of KZB
equations is known to be related to the quantum (and classical) CM
models by the Matsuo-Cherednik construction \cite{Matsuo,Cherednik2}
(see also \cite{MTV}). Recently relations between CM (and
Ruijsenaars-Schneider (RS)) models to quantum spin chains were
actively investigated \cite{Anton,GZZ}.

\vskip2mm

{\bf 2. $R$-matrix valued Lax pairs}

\vskip2mm

The Fay type identities (\ref{iq702}) for the quantum $R$-matrices
allows to suggest extended version of the Krichever's ansatz for CM
Lax pairs with spectral parameter \cite{Krich1}. Consider the
following block matrix Lax operator
  \beq\label{iq301}
 \begin{array}{c}
  \displaystyle{
\mathcal L=\sum\limits_{a,b=1}^{\ti N} \ti{\mathrm E}_{ab}\otimes
\mathcal L_{ab}
 }
 \end{array}
 \eq
where $\ti{\mathrm E}_{ab}$ is the standard basis of ${\rm gl}_{\ti
N}$ and
 \beq\label{iq302}
   \begin{array}{|c|}
  \hline\\
 \displaystyle{
\mathcal L_{ab}=\delta_{ab}p_a\,1_a\otimes
1_b+\nu(1-\delta_{ab})R_{ab}^\hbar\,,\ \ \
R_{ab}^\hbar=R_{ab}^\hbar(z_a-z_b)\,.
  }
   \\ \ \\
 \hline
 \end{array}
 \eq
When $N=1$ the ${\rm gl}_N$ $R$-matrix reduces to its scalar
analogue -- function $\phi(z,\hbar)$ and we reproduce the answer
from \cite{Krich1} for $\tilde N$-body CM system.
 Notice that the Planck constant of ${\rm gl}_N$
$R$-matrix plays here the role of the spectral parameter for ${\rm
gl}_{\ti N}$ CM model. The corresponding $M$-operator is given in
(\ref{q320}). The Lax equation $\p_t\mathcal L=[\mathcal L,\mathcal
M]$ is equivalent to dynamics of $\ti N$ CM particles
 \beq\label{iq326}
 \begin{array}{c}
  \displaystyle{
\ddot z_a=N^2\nu^2\sum\limits_{b\neq a}\wp'(z_a-z_b)\,.
 }
 \end{array}
 \eq
In the same way the monodromy preserving equation
$
\p_\tau\mathcal L-\p_\hbar \mathcal M=[\mathcal L,\mathcal M]
$
leads to the Painlev\'e equations
 \beq\label{q036}
 \begin{array}{c}
  \displaystyle{
\p_\tau^2 z_a=N^2\nu^2\sum\limits_{b\neq a}\wp'(z_a-z_b)\,.
 }
 \end{array}
 \eq
The corresponding linear problem has the form
 \beq\label{q037}
 \begin{array}{c}
  \displaystyle{
(\p_\hbar+\mathcal L)\Psi=0\,.
 }
 \end{array}
 \eq
Let us also mention that the linear problem for the quantum version
of CM model
  \beq\label{q038}
 \begin{array}{c}
  \displaystyle{
\hat{\mathcal L}\Psi=\Psi\Lambda\,,\ \ \ \hat{\mathcal
L}_{ab}=\delta_{ab}\,\p_{z_a}\,1_a\otimes
1_b+\nu(1-\delta_{ab})R_{ab}^\hbar
 }
 \end{array}
 \eq
resembles very much the KZ connections from the first line of
(\ref{iq140}). Equation (\ref{q038}) (or (\ref{q037}) with
$\hat{\mathcal L}$) generalizes the first line of (\ref{iq140}) in
the same way as the Lax pairs with spectral parameter generalize
those without it. We hope to clarify exact relations between
$R$-matrix valued linear problems and KZB equations in our future
papers.

Choosing elliptic, trigonometric or the rational $R$-matrix we
describe the CM models similarly to ${\rm gl}_1$ case \cite{Krich1}.
Notice that the ${\rm gl}_N$ $R$-matrix itself describes ${\rm
gl}_N$ integrable systems such as integrable tops which are gauge
equivalent to CM or RS models. Here we use ${\rm gl}_N$ $R$-matrices
as auxiliary spaces for derivation of ${\rm gl}_{\ti N}$ models.
The next natural step is to get similar result for the
Ruijsenaars-Schneider (quantum) model. In this case we deal with two
Planck constants. Our general idea is that the both Planck constants
can play different roles, i.e. each of the constants can be either
the spectral parameter in a "classical-quantum" ${\rm gl}_{\ti N}$
system (of (\ref{iq302}) type) or the Planck constant in a quantum
${\rm gl}_N$ system or the relativistic deformation parameter in a
classical relativistic ${\rm gl}_N$ model (see
\cite{LOZ7})\footnote{Let us also remark that in \cite{LOSZ} we have
already found an $R$-matrix intermediate between the Belavin's and
the Felders' one. Her we use a different description. Presumably,
the interrelation between different descriptions is given by the
Fourier-Mukai type transformation.}.
We hope that this can shed light on numerous dualities in integrable
systems mentioned in \cite{Nekrasov}, \cite{MMRZZ}, \cite{ZZ},
\cite{Gaiotto}.

\vskip3mm

{\small

\noindent {\bf Acknowledgments.} The work was supported by RFBR
grants 12-02-00594 (A.L. and M.O.) and 12-01-00482 (A.Z.). The work
of A.L. was partially supported by AG Laboratory GU-HSE, RF
government grant, ag. 11 11.G34.31.0023 and by the Simons
Foundation. The work of A.Z. was partially supported by the D.
Zimin's fund "Dynasty", by the Program of RAS "Basic Problems of the
Nonlinear Dynamics in Mathematical and Physical Sciences"  $\Pi$19
and by grant RSCF 14-50-00005.

}

\section{From integrable tops to KZB equations}
\setcounter{equation}{0}

In this section we describe the sequence of steps which leads to the
KZB equations \cite{KZB} starting from integrable tops. As it was
mentioned above, our consideration is independent on the choice of
particular top model. The basic element is the underlying quantum
$R$-matrix \cite{LOZ7}.

First, we briefly recall the structures underlying integrable tops
and proceed to the non-autonomous dynamics. It is described by the
monodromy preserving equations. In the same way the Schlesinger
system is originated  from the corresponding Gaudin model. At last,
the KZB equations arise from the quantization of the Schlesinger
system \cite{Resh,LO2,Korotkin}.


\subsection{Integrable tops}
In \cite{LOZ7} we defined the relativistic integrable top by means
of the quantum $R$-matrix. The ${\rm gl}_N$ Lax matrix is given by
 \beq\label{q101}
 \begin{array}{c}
  \displaystyle{
L^\eta(z,S)=\tr_2(R_{12}^\eta(z) S_2)\,,\ \ \
S=\res\limits_{z=0}L^\eta(z,S)\,,
  }
 \end{array}
 \eq
where $S=\sum\limits_{i,j=1}^N {\mathrm E}_{ij}S_{ij}$  is the ${\rm
gl}_N$-valued dynamical variable\footnote{$\{{\mathrm E}_{ij},\
i,j=1...N\}$ is the standard basis in the fundamental representation
of ${\rm gl}_N$: $({\mathrm E}_{ij})_{kl}=\delta_{ik}\delta_{jl}$.},
and $R_{12}^\eta(z)$ is the corresponding quantum non-dynamical
$R$-matrix. It satisfies the quantum Yang-Baxter equation
(\ref{q102}). The non-relativistic limit ($\eta\to 0$)
 \beq\label{q103}
 \begin{array}{c}
  \displaystyle{
L^\eta(z,S)= \eta^{-1}\,\frac{\tr S}{N}\,1_{N\times
 N}+L(z,S)+\eta\,{\mathcal M}(z,S)+O(\eta^2)
 }
 \end{array}
 \eq
is related to the classical limit ($\hbar\to 0$) (\ref{q104})
via (\ref{q101}):
 \beq\label{q1041}
 \begin{array}{c}
  \displaystyle{
L(z,S)=\tr_2\left(r_{12}(z)S_2\right)\,, \ \ \
S=\res\limits_{z=0}L(z,S)\,,
 }
 \end{array}
 \eq
  \beq\label{q1042}
 \begin{array}{c}
  \displaystyle{
\mathcal M(z,S)=\tr_2\left(m_{12}(z)S_2\right)\,.
 }
 \end{array}
 \eq
The quantity $r_{12}(z)$ in (\ref{q104}), (\ref{q1041}) is the
classical $r$-matrix. It is skew-symmetric (\ref{q043})
  \beq\label{q1043}
 \begin{array}{c}
  \displaystyle{
r_{12}(z)=-r_{21}(-z)
 }
 \end{array}
 \eq
and satisfies the classical Yang-Baxter equation:
 \beq\label{q105}
 \begin{array}{c}
  \displaystyle{
[r_{12}(z-w),r_{13}(z)]+[r_{12}(z-w),r_{23}(w)]+[r_{13}(z),r_{23}(w)]=0\,.
 }
 \end{array}
 \eq
As it was mentioned in \cite{LOZ7} the matrices (\ref{q1041}),
(\ref{q1042}) appear to be the Lax pair of the non-relativistic top.
It means that the Lax equation
 \beq\label{q106}
 \begin{array}{c}
  \displaystyle{
\p_t L(z,S)=[L(z,S),\mathcal M(z,S)]
 }
 \end{array}
 \eq
is equivalent to equations of motion
 \beq\label{q107}
 \begin{array}{c}
  \displaystyle{
\p_t S=[S,J(S)]\,,
 }
 \end{array}
 \eq
where the inverse inertia tensor is given by the linear functional
 \beq\label{q108}
 \begin{array}{c}
  \displaystyle{
J(S)=\mathcal M(0,S)\,.
 }
 \end{array}
 \eq
The equations (\ref{q107}) are Hamiltonian with the Hamiltonian
function
 \beq\label{q109}
 \begin{array}{c}
  \displaystyle{
{\mathcal H}^{\hbox{\tiny{top}}}(S)=\frac{1}{2}\,\tr(S\, J(S))
 }
 \end{array}
 \eq
and the Poisson-Lie brackets on ${\rm gl}_N^*$
  \beq\label{q110}
 \begin{array}{c}
  \displaystyle{
\{S_1,S_2\}=[S_2,P_{12}]
  }
 \end{array}
 \eq
 or $\{S_{ij},S_{kl}\}=\delta_{il}S_{kj}-\delta_{kj}S_{il}$.

\subsection{Painlev\'e--Calogero correspondence and non-autonomous tops}

The (classical) Painlev\'e--Calogero correspondence was suggested in
\cite{LO}. It claims that the (Krichever's) Lax pair of the elliptic
Calogero-Moser model can be also used for the monodromy preserving
equations, which describe the higher rank Painlev\'e equations in
the elliptic form.

Let us formulate here the Painlev\'e--Calogero correspondence in the
form of the quantum non-dynamical $R$-matrix property.
 \begin{defi}
Suppose that the quantum $R$-matrix entering (\ref{q101}) depends on
some additional parameter $\tau$:
$R^{\hbar,\tau}(z)=R(z,\hbar,\tau)$. We say that the $R$-matrix
 satisfies the property of the
"Painlev\'e--Calogero cor\-respondence" if the following relation
holds\,\footnote{Notice that the definition depends on the gauge
choice.}:
  \beq\label{q1117}
 \begin{array}{c}
  \displaystyle{
\p_\tau R^{\hbar,\tau}(z)=\p_z\p_\hbar R^{\hbar,\tau}(z)\,.
  }
 \end{array}
 \eq
 \end{defi}
Plugging the expansion (\ref{q104}) into  (\ref{q1117}) we get a set
of relations. The first non-trivial is
  \beq\label{q111}
 \begin{array}{c}
  \displaystyle{
\p_\tau r^\tau_{12}(z)=\p_z m^\tau_{12}(z)\,,
  }
 \end{array}
 \eq
where $r^\tau_{12}(z)=r_{12}(z,\tau)$ is the classical $r$-matrix.
%
%
An example of the $R$-matrix with this property is given by the
Baxter-Belavin's one \cite{Baxter} (see Appendix B). The parameter
$\tau$ in this example equals $\tau^{ell}/2\pi\imath$, where
$\tau^{ell}$ is the module of the underlying elliptic curve, and the
property (\ref{q111}) is due to the heat equation for the
theta-functions
  \beq\label{q112}
 \begin{array}{c}
  \displaystyle{
2\p_\tau\vth(z|\tau)=\p_z^2\vth(z|\tau)\,.
  }
 \end{array}
 \eq
From (\ref{q111}) and (\ref{q1041})-(\ref{q1042}) it follows that
  \beq\label{q113}
 \begin{array}{c}
  \displaystyle{
\frac{\p}{\p \tau}\, L^\tau(z,S)=\frac{\p}{\p z}\, \mathcal
M^\tau(z,S)\,,
  }
 \end{array}
 \eq
where $L^\tau(z,S)=L(z,S,\tau)$, $\mathcal M^\tau(z,S)=\mathcal
M(z,S,\tau)$. Therefore, we can define the monodromy preserving
equations in time $\tau$
  \beq\label{q114}
 \begin{array}{c}
  \displaystyle{
d_\tau L^\tau(z,S)-\p_z\mathcal M^\tau(z,S)=[L^\tau(z,S),\mathcal
M^\tau(z,S)]\,,\ \ \ S=S(\tau)
  }
 \end{array}
 \eq
($d_\tau=\frac{d}{d\tau}$) as the non-autonomous version of the
integrable top's equations of motion (\ref{q107})\footnote{These
models are no more integrable but can be treated as alternative
description of (higher) Painlev\'e equations. See \cite{LOZ2} for
the example of Painlev\'e VI.}:
 \beq\label{q115}
 \begin{array}{c}
  \displaystyle{
\p_\tau S=[S,J^\tau(S)]\,.
 }
 \end{array}
 \eq
Indeed, the total derivative $d_\tau L^\tau(z,S)$ contains both --
the partial derivatives by explicit and implicit dependence on
$\tau$:
 \beq\label{q116}
 \begin{array}{c}
  \displaystyle{
d_\tau L^\tau(z,S(\tau))=d_\tau
\tr_2(r_{12}^\tau(z)S_2)=\tr_2\Big((\p_\tau
r_{12}^\tau(z))\,S_2\Big)+\tr_2\Big( r_{12}^\tau(z)\,(\p_\tau
S_2)\Big)\,.
 }
 \end{array}
 \eq
The first term is cancelled by $\p_z\mathcal M^\tau(z,S)$
(\ref{q113}), and we get the same result as in (\ref{q107})
following from the Lax equations (\ref{q106}). But this time it
contains explicit dependence on $\tau$ via
 \beq\label{q117}
 \begin{array}{c}
  \displaystyle{
J^\tau(S)=\mathcal M^\tau(0,S)\,.
 }
 \end{array}
 \eq
Similarly to the autonomous case this system is Hamiltonian (see
(\ref{q109})) with
 \beq\label{q118}
 \begin{array}{c}
  \displaystyle{
{\mathcal H}^\tau(S)=\frac{1}{2}\,\tr(S\, J^\tau (S))
 }
 \end{array}
 \eq
and the Poisson brackets are given by (\ref{q110}).

Let us keep the notation $\frac{\p}{\p \tau}$ (but not $\p_\tau$)
for the partial derivative by only explicit dependence on $\tau$,
i.e.
 \beq\label{q1181}
 \begin{array}{c}
  \displaystyle{
\frac{\p}{\p \tau} L^\tau(z,S(\tau))=\tr_2\Big((\p_\tau
r_{12}^\tau(z))\,S_2(\tau)\Big)\,.
 }
 \end{array}
 \eq

\subsection{Gaudin models}

The phase space of the Gaudin model \cite{Gaudin}
is the direct product of $n$ coadjoint orbits, i.e. $\ti N$ copies
of $S$: $S^a\in {\rm gl}_N$, $a=1,...,\ti N$ with some fixed
eigenvalues. Its Poisson structure
 \beq\label{q119}
   \displaystyle{
 \{S_1^a,S_2^b\}=\delta^{ab}\,[S_2^a,P_{12}]
 }
 \eq
 is the direct sum of (\ref{q110}).
The Lax matrix has $n$ simple poles at $\{z_a,\ a=1,...,\ti N\}$
with residues $S^a$. It is given in terms of the top Lax matrix
(\ref{q1041}):
 \beq\label{q120}
 \begin{array}{c}
  \displaystyle{
  L^{\hbox{\tiny{G}}}(z)=\sum\limits_{a=1}^{\ti N} L^\tau(z-z_a, S^a)=
  \sum\limits_{a=1}^{\ti N} \tr_2\Big(r_{12}^\tau(z-z_a) S_2^a\Big)\,.
 }
 \end{array}
 \eq
Here we imply the existence of the deformation parameter $\tau$
(\ref{q112})-(\ref{q118}) from the very beginning in order not to
repeat (almost) the same notations with $\tau$ and without $\tau$ as
we made for the top and its non-autonomous version.

We consider the flows corresponding to Hamiltonians
 \beq\label{q121}
 \begin{array}{c}
  \displaystyle{
  h_a=-\sum\limits_{c\neq a}^{\ti N}
\tr \left( S^a\,L^\tau(z_a-z_c, S^c)\right)=-\sum\limits_{c\neq
a}^{\ti N}\tr_{12}\Big(r_{12}^\tau(z_a-z_c) S^a_1  S^c_2\Big)
 }
 \end{array}
 \eq
for $a=1,...,\ti N$ and
 \beq\label{q122}
 \begin{array}{c}
  \displaystyle{
{\mathcal H}_0=\frac{1}{2}\sum\limits_{b,c\,=1}^{\ti
N}\tr\left(S^b\,{ \mathcal
M^\tau}(z_b-z_c,S^c)\right)=\frac{1}{2}\sum\limits_{b,c\,=1}^{\ti N}
\tr_{12}\Big(m_{12}^\tau(z_a-z_c) S^b_1  S^c_2\Big)\,.
 }
 \end{array}
 \eq
Notice that the terms coming from $b=c$ in (\ref{q122}) are the top
Hamiltonians ${\mathcal H}^\tau(S^c)$ (\ref{q118}). The functions
(\ref{q121})-(\ref{q122}) Poisson commute because (\ref{q119}) is
equivalent to the classical exchange relations
 \beq\label{q123}
 \begin{array}{c}
  \displaystyle{
\{L_1^{\hbox{\tiny{G}}}(z),L_2^{\hbox{\tiny{G}}}(w)\}=[L_1^{\hbox{\tiny{G}}}(z)+L_2^{\hbox{\tiny{G}}}(w),r^\tau_{12}(z-w)]\,.
  }
 \end{array}
 \eq
\vskip1mm

\noindent The dynamics generated by  (\ref{q121})-(\ref{q122})
 \beq\label{q124}
 \left\{\begin{array}{l}
  \displaystyle{
 \p_{t_a} S^b=-[S^b,L^\tau(z_a-z_b,S^a)]\,,\ \ b\neq a}
\\ \ \\
  \displaystyle{
\p_{t_a} S^a=\sum\limits_{c\neq a}^n\, [S^a,L^\tau(z_c-z_a,S^c)]
 }
 \end{array}\right.
 \eq
for $a=1,...,\ti N$ and
 \beq\label{q125}
 \begin{array}{c}
  \displaystyle{
 \p_{t_0}S^a=[S^a,J^\tau (S^a)]+\sum\limits_{c\neq a}\,
 [S^a,\mathcal M^\tau(z_a-z_c,S^c)]
 }
 \end{array}
 \eq
possesses the Lax representations
 \beq\label{q126}
 \begin{array}{c}
  \displaystyle{
 \p_{t_d} L^{\hbox{\tiny{G}}}(z)=[L^{\hbox{\tiny{G}}}(z),M^{\hbox{\tiny{G}},\,d}]\,,\ \ d=0,...,\ti N
 }
 \end{array}
 \eq
where
 \beq\label{q127}
 \begin{array}{c}
  \displaystyle{
 M^{\hbox{\tiny{G}},\,a}(z)=-L^\tau(z-z_a,S^a)\,,\ \ a=1,...,\ti N
 }
 \end{array}
 \eq
and
 \beq\label{q128}
 \begin{array}{c}
  \displaystyle{
 M^{\hbox{\tiny{G}},\,0}(z)=\sum\limits_{c=0}^{\ti N} \mathcal
 M^\tau(z-z_c,S^c)\,.
 }
 \end{array}
 \eq

\subsection{Schlesinger systems}

Similarly to the description of Painlev\'e equation in the form of
non-autonomous tops let us also represent the Schlesinger
 system \cite{Schlesinger} as the non-autonomous Gaudin model.

 First, it follows from
(\ref{q120}) and (\ref{q127}) that
 \beq\label{q130}
 \begin{array}{c}
  \displaystyle{
 \frac{\p}{\p z_a}\,  L^{\hbox{\tiny{G}}}(z)=\frac{\p}{\p z}\,  M^{\hbox{\tiny{G}},\,a}(z)\,.
 }
 \end{array}
 \eq
Secondly, it follows from (\ref{q120}), (\ref{q128}) and
(\ref{q113}) that\footnote{In (\ref{q130}) and (\ref{q131}) the
partial
 derivatives are taken with respect to explicit dependence on $\tau$ or $z_a$
 (\ref{q1181}).}
 \beq\label{q131}
 \begin{array}{c}
  \displaystyle{
 \frac{\p}{\p \tau} \, L^{\hbox{\tiny{G}}}(z)=\frac{\p}{\p z}\,  M^{\hbox{\tiny{G}},\,0}(z)\,.
 }
 \end{array}
 \eq
Therefore, the monodromy preserving equations (or compatibility
conditions for isomonodromic deformations)
 \beq\label{q132}
 \begin{array}{c}
  \displaystyle{
 \p_{z_a} L^{\hbox{\tiny{G}}}(z) -\p_z M^{\hbox{\tiny{G}},\,a}(z)=[L^{\hbox{\tiny{G}}}(z), M^{\hbox{\tiny{G}},\,a}(z)]
 }
 \end{array}
 \eq
and
 \beq\label{q133}
 \begin{array}{c}
  \displaystyle{
 \p_\tau L^{\hbox{\tiny{G}}}(z) -\p_z M^{\hbox{\tiny{G}},\,0}(z)=[L^{\hbox{\tiny{G}}}(z), M^{\hbox{\tiny{G}},\,0}(z)]
 }
 \end{array}
 \eq
generate dynamics in time variables $z_a$ and $\tau$. They have form
form of non-autonomous versions of the Gaudin's one
(\ref{q124})-(\ref{q125}):
 \beq\label{q134}
 \left\{\begin{array}{l}
  \displaystyle{
 \p_{z_a} S^b=-[S^b,L^\tau(z_a-z_b,S^a)]\,,\ \ b\neq a}
\\ \ \\
  \displaystyle{
\p_{z_a} S^a=\sum\limits_{c\neq a}^{\ti N}\,
[S^a,L^\tau(z_c-z_a,S^c)]
 }
 \end{array}\right.
 \eq
for $a=1,...,\ti N$ and
 \beq\label{q135}
 \begin{array}{c}
  \displaystyle{
 \p_\tau S^a=[S^a,J^\tau (S^a)]+\sum\limits_{c\neq a}\,
 [S^a,\mathcal M^\tau(z_a-z_c,S^c)]\,.
 }
 \end{array}
 \eq
The Hamiltonians (\ref{q121})-(\ref{q122}) and the Poisson structure
(\ref{q119}) are of the same form\footnote{The elliptic case was
considered in \cite{LO2,Korotkin,CLOZ1,LOZ5}.}.

\subsection{KZB equations}

The relation between KZB equations and the quantum monodromy
preserving equations was described in \cite{Resh} (see also
\cite{LO2,Korotkin}).
Let us formulate it using notations of (\ref{q104}) with the
$\tau$-deformation satisfying (\ref{q111}).   The KZB equations have
form:
 \beq\label{q140}
 \left\{
 \begin{array}{l}
  \displaystyle{
\hat\nabla_a \psi=0\,,
 }
 \\
   \displaystyle{
\hat\nabla_\tau \psi=0\,,
 }
 \end{array}
 \right.
 \eq
where
 \beq\label{q144}
 \begin{array}{c}
  \displaystyle{
\nabla_a=\p_{z_a}+\sum\limits_{c\neq a} {\mathfrak
r}^\tau_{ac}(z_a-z_c)\,,
 }
 \end{array}
 \eq
 \beq\label{q145}
 \begin{array}{c}
  \displaystyle{
\nabla_\tau=\p_\tau+\frac{1}{2}\sum\limits_{b,c} {\mathfrak
m}^\tau_{bc}(z_b-z_c)\,.
 }
 \end{array}
 \eq
Here $\mathfrak r^\tau_{ac}$ and $\mathfrak m^\tau_{ac}$ are the
operators acting by $a$-th and $c$-th components of ${\rm U}({\rm
gl}_N)^{\otimes \ti N}$ (the tensor product of $\ti N$ copies of the
universal enveloping algebra). Recall that in classical integrable
systems (as well as in the Schlesinger systems) we used the
fundamental representation $\rho_N$ of ${\rm gl}_N$ (see e.g.
(\ref{q1041})-(\ref{q1042})):
 \beq\label{q1451}
 \begin{array}{c}
  \displaystyle{
r^\tau_{12}(z)=\rho_N({\mathfrak
r}^\tau_{12}(z))=\sum\limits_{i,j,k,l}r^\tau_{ij,kl}\,\mathrm
E_{ij}\otimes \mathrm E_{kl}\,,
 }
 \\ \ \\
  \displaystyle{
m^\tau_{12}(z)=\rho_N({\mathfrak
m}^\tau_{12}(z))=\sum\limits_{i,j,k,l}m^\tau_{ij,kl}\,\mathrm
E_{ij}\otimes \mathrm E_{kl}\,,
 }
 \end{array}
 \eq
The algebra ${\rm U}({\rm gl}_N)^{\otimes \ti N}$ can be considered
as a quantization of the classical phase space with the Poisson
structure (\ref{q119}). Indeed, let
 \beq\label{q137}
 \begin{array}{c}
  \displaystyle{
S^a\ \to\ {\hat S}^a:\ \ {\hat S}^a_{ij}:={\bfe}_{ji}^a\,,
 }
 \end{array}
 \eq
 where $\{{\bfe}_{ij}^a\}$: $[\bfe^a_{ij},\bfe^a_{kl}]=\delta^{ab}(\bfe^a_{il}\delta_{kj}-\bfe^a_{kj}\delta_{il})$
  is the standard basis in the $a$-th
 component of ${\rm U}({\rm gl}_N)^{\otimes \ti N}$.
  In this notation
 \beq\label{q1371}
 \begin{array}{c}
  \displaystyle{
\mathfrak
r^\tau_{ab}=\sum\limits_{i,j,k,l}r^\tau_{ij,kl}(z_a-z_b)\,\bfe_{ij}^a\bfe_{kl}^b=\sum\limits_{i,j,k,l}r^\tau_{ij,kl}(z_a-z_b)\,{\hat
S}^a_{ji}{\hat S}^b_{lk}\,,
 }
 \end{array}
 \eq
  \beq\label{q1372}
 \begin{array}{c}
  \displaystyle{
\mathfrak
m^\tau_{ab}=\sum\limits_{i,j,k,l}m^\tau_{ij,kl}(z_a-z_b)\,\bfe_{ij}^a\bfe_{kl}^b=\sum\limits_{i,j,k,l}m^\tau_{ij,kl}(z_a-z_b)\,{\hat
S}^a_{ji}{\hat S}^b_{lk}\,.
 }
 \end{array}
 \eq
 The fundamental representation is given by
 $\rho_N({\bfe}_{ij}^a)=1\otimes ...\otimes 1 \otimes {\mathrm E}_{ij}\otimes 1\otimes ...\otimes
 1$, where $\left({\mathrm
 E}_{ij}\right)_{kl}=\delta_{ik}\delta_{jl}$ is on the $a$-th
  place. Then $r$-matrix is an operator acting on the $a$-th and
$b$-th components of an element of the tensor product $V^{\otimes
\ti N}$. The operator is represented by matrix of $N^{\ti N}\times
N^{\ti N}$ size because it also contains (as factors) the product of
identity operators for the rest of components
$\bigotimes\limits_{c\neq a,b}1_c$.
 The residue of $r$-matrix is (up
to factor $N$ in (\ref{q1495})) the permutation operator replacing
$a$-th and $b$-th components of an element of the tensor product
$V^{\otimes \ti N}$ to which $\psi$  belongs.

 Then
  \beq\label{q138}
 \begin{array}{c}
  \displaystyle{
[{\hat S}_0^a,{\hat S}_{0'}^b]=\delta^{ab}\,[{\hat
S}_{0'}^a,P_{00'}]\,,\ \ \hat S^a=\sum\limits_{i,j=1}^N\hat
S^a_{ij}\,\rho_N({\bfe}^a_{ij})
 }
 \end{array}
 \eq
or $[{\hat S}_{ij}^a,{\hat S}_{kl}^b]=\delta^{ab}\,\left( {\hat
S}_{kj}^a \delta_{il}-{\hat S}_{il}^a \delta_{kj} \right)$. The
indices $0,0'$ in (\ref{q138}) are the  notations for the components
of $\left(\rho_N({\rm U}({\rm gl}_N)^{\otimes \ti
N})\right)^{\otimes 2}$ -- tensor product of auxiliary spaces.
To quantize the Hamiltonian (\ref{q122}) we also need to fix the
ordering. Consider the symmetric (Weyl) ordering
  \beq\label{q139}
 \begin{array}{c}
  \displaystyle{
\widehat{{S}_{ij}^a {S}_{kl}^b}=\frac{1}{2}\Big({\hat S}_{ij}^a
{\hat S}_{kl}^b+{\hat S}_{kl}^b {\hat S}_{ij}^a\Big).
 }
 \end{array}
 \eq
Then the KZB connections (\ref{q144})-(\ref{q145}) are written in
terms of the quantum versions of the classical Hamiltonians $h_a$
and ${\mathcal H}_0$ (\ref{q121})-(\ref{q122}):
   \beq\label{q143}
 \begin{array}{c}
  \displaystyle{
\hat\nabla_a=\p_{z_a}-\hat h_a\,,\ \ \ \hat\nabla_\tau=\p_\tau+\hat
{\mathcal H}_0\,.
 }
 \end{array}
 \eq
In the same time the KZB equations (\ref{q140}) acquire the form of
the non-stationary Schr\"odinger equations in times $z_1,...,z_{\ti
N}$ and $\tau$.

The compatibility conditions of KZB equations (\ref{q140})
   \beq\label{q141}
 \begin{array}{c}
  \displaystyle{
[\hat\nabla_a,\hat\nabla_b]=0
 }
 \end{array}
 \eq
   \beq\label{q142}
 \begin{array}{c}
  \displaystyle{
[\hat\nabla_a,\hat\nabla_\tau]=0
 }
 \end{array}
 \eq
are fulfilled identically\footnote{This statement was verified
directly in different cases. See
 \cite{Etingof,Korotkin} for elliptic examples.}. The first one
 (\ref{q141}) follows from the classical Yang-Baxter equation
   \beq\label{q146}
 \begin{array}{c}
  \displaystyle{
[{\mathfrak r}_{ab},{\mathfrak r}_{bc}]+[{\mathfrak
r}_{bc},{\mathfrak r}_{ac}]+[{\mathfrak r}_{ab},{\mathfrak
r}_{ac}]=0\,,
 }
 \end{array}
 \eq
where ${\mathfrak r}_{ab}={\mathfrak r}_{ab}^\tau(z_a-z_b)$. The set
of identities underlying (\ref{q141}) consists of the property
(\ref{q111})
   \beq\label{q1461}
 \begin{array}{c}
  \displaystyle{
\p_\tau{\mathfrak r}_{ab}=\p_{z_a}{\mathfrak m}_{ab}\,,
 }
 \end{array}
 \eq
where ${\mathfrak m}_{ab}={\mathfrak m}_{ab}^\tau(z_a-z_b)$ and
   \beq\label{q147}
 \begin{array}{c}
  \displaystyle{
\frac12\,[{\mathfrak r}_{ab},{\mathfrak m}_{aa}+{\mathfrak
m}_{bb}]+[{\mathfrak r}_{ab},{\mathfrak m}_{ab}]=0\,,
 }
 \end{array}
 \eq
   \beq\label{q148}
 \begin{array}{c}
  \displaystyle{
[{\mathfrak r}_{ab},{\mathfrak m}_{bc}]+[{\mathfrak
r}_{ab},{\mathfrak m}_{ac}]+[{\mathfrak r}_{ac},{\mathfrak
m}_{ab}]+[{\mathfrak r}_{ac},{\mathfrak m}_{bc}]=0\,.
 }
 \end{array}
 \eq
%

%
\begin{rem}
 One can get more identities relating $r_{ab}$ and $m_{ab}$ and higher order terms of expansion (\ref{q104})
 from the Yang-Baxter
equation (\ref{q102})
$R_{ab}^{\hbar,\tau}R_{ac}^{\hbar,\tau}R_{bc}^{\hbar,\tau}=R_{bc}^{\hbar,\tau}R_{ac}^{\hbar,\tau}R_{ab}^{\hbar,\tau}$.
The first non-trivial identity is (\ref{q146}). The next one is
   \beq\label{q149}
 \begin{array}{c}
  \displaystyle{
[r_{ab},m_{ac}]+[m_{ab},r_{ac}]+[r_{ab},m_{bc}]+[m_{ab},r_{bc}]+[r_{ac},m_{bc}]+[m_{ac},r_{bc}]+
 }
 \\ \ \\
  \displaystyle{
+r_{ab}r_{ac}r_{bc}-r_{bc}r_{ac}r_{ab}=0\,,
 }
 \end{array}
 \eq
where ${r}_{ab}={r}_{ab}^\tau(z_a-z_b)$,
${m}_{ab}={m}_{ab}^\tau(z_a-z_b)$.
 \end{rem}

\section{Rational non-autonomous tops and KZB equations}
\setcounter{equation}{0}

 The rational top was first studied for small rank cases in
 \cite{Smirnov1} by degenerating the elliptic Lax matrix \cite{LOZ}. Later it was
 constructed for ${\rm gl}_N$ case using its relation to the rational Calogero-Moser
 model \cite{AASZ}. The idea was to compute the classical (skew-symmetric non-dynamical) $r$-matrix
 as follows:
   \beq\label{q151}
 \begin{array}{c}
  \displaystyle{
r_{12}(z)=\frac{\p L_1(z,S)}{\p S_2}\,,\ \
S=\res\limits_{z=0}L(z)\,.
 }
 \end{array}
 \eq
%
In \cite{LOZ7} this relation was extended to the quantum $R$-matrix
by proceeding to the relativistic top:
   \beq\label{q153}
 \begin{array}{c}
  \displaystyle{
R_{12}^\hbar(z)=\frac{\p L_1^\hbar(z,S)}{\p S_2}\,,\ \ \
S=\res\limits_{z=0}L^\hbar(z)\,,
 }
 \end{array}
 \eq
where the classical Lax matrix $L^\hbar(z)$ depends on the constant
$\hbar$ playing the role of the relativistic deformation parameter.
The Lax matrix was found using its relation to the
Ruijsenaars-Schneider (RS) model. In the spinless case the gauge
transformation relating two models
   \beq\label{q154}
 \begin{array}{c}
  \displaystyle{
L^\eta(z, S)=g(z)L^{\hbox{\tiny{RS}}}(z,\eta)g^{-1}(z)
 }
 \end{array}
 \eq
can be written explicitly in terms of the RS particles coordinates
$q_j$: $g(z,q)=\Xi(z,q)D^{-1}$, where\footnote{The explicit from of
$L^{\hbox{\tiny{RS}}}(z,\eta)$ as well  as diagonal matrix
$D_{ij}=\delta_{ij}\prod\limits_{k\neq i}(q_i-q_k)$ is not used in
what follows.}
   \beq\label{q155}
 \begin{array}{c}
  \displaystyle{
\Xi(z,q)=(z+q_j)^{\varrho(i)}\,,
 }
 \\ \ \\
   \displaystyle{
\varrho(i)=i-1\ \ \hbox{for}\ \ i\leq N-1;\ \ \ \varrho(N)=N\,.
 }
 \end{array}
 \eq

\subsection{$\tau$-deformation of quantum rational $R$-matrix}
 Our aim is to construct
$\tau$-dependent $R$-matrix satisfying the Painlev\'e-Calogero
property (\ref{q1117}) starting from the $\tau$-independent one
(\ref{q153}). The answer follows from (\ref{q159}) (see below). It
appears that the deformation of the Yang's rational $R$-matrix
suggested in \cite{LOZ7} admits this kind of deformation similarly
to the elliptic case. The idea is to deform first $\Xi(z)$
(\ref{q155}). Let us find $\Xi(z,q|\,\tau)$ satisfying the heat
equation
   \beq\label{q156}
 \begin{array}{c}
  \displaystyle{
2\p_\tau \Xi(z|\,\tau)=\p_z^2\,\Xi(z|\,\tau)
 }
 \end{array}
 \eq
with the boundary condition
  \beq\label{q157}
 \begin{array}{c}
  \displaystyle{
\Xi(z|\,0)=\Xi(z)\,.
 }
 \end{array}
 \eq
 Then the $R$-matrices (\ref{q151}), (\ref{q153}) constructed by
means of $\Xi(z|\,\tau)$ satisfy the property
(\ref{q1117})\footnote{It can be also proved directly by using
explicit answer for the quantum $R$-matrix \cite{LOZ7}.}. The
solution of (\ref{q156})-(\ref{q157}) is given by
  \beq\label{q158}
 \begin{array}{c}
  \displaystyle{
\Xi(z|\,\tau)=\exp\left(\,\frac{\tau}{2}\,\p_z^2\,\right) \Xi(z)
 }
 \end{array}
 \eq
or
  \beq\label{q159}
 \begin{array}{c}
  \displaystyle{
\Xi(z|\,\tau)=\exp\left(\,\frac{\tau}{2}\, T\,\right) \Xi(z)\,,
 }
 \end{array}
 \eq
where $T$ is the nilpotent operator representing the action of
$\p_z^2$ on the $N$-dimensional column-vector
$(1,z,z^2,...,z^{N-2},z^N)^T$. It is $N\times N$ matrix with
elements
  \beq\label{q160}
 T_{ij}=\left\{
 \begin{array}{l}
  \displaystyle{
j(j+1)\delta_{i-2,j}\,,\ i<N\,,
 }
 \\
  \displaystyle{
j(j+1)\delta_{i-1,j}\,,\ i=N\,.
 }
 \end{array}\right.
 \eq
For example, for $N=2,3,4$ we have:
  \beq\label{q161}
 \begin{array}{c}
  \displaystyle{
 T_{N=2}=\left( \begin {array}{cc}
0&0\\\noalign{\medskip}2&0\end {array}
 \right)\,,\ \ \ \ \
 T_{N=3}=\left( \begin {array}{ccc} 0&0&0\\\noalign{}0&0&0
\\\noalign{}0&6&0\end {array} \right)\,,\ \ \ \ \
 T_{N=4}= \left( \begin {array}{cccc} 0&0&0&0\\\noalign{}0&0&0&0
\\\noalign{}2&0&0&0\\\noalign{}0&0&12&0\end {array}
 \right)\,.

 }
 \end{array}
 \eq
 Denote
  \beq\label{q1611}
 \begin{array}{c}
  \displaystyle{
\mathcal T:= \exp\left(\,\frac{\tau}{2}\, T\,\right)\,,
  }
 \end{array}
 \eq
i.e. $\Xi(z|\,\tau)=\mathcal T\,\Xi(z|\,0)$. Then for $N=2,3,4$ the
operator $\mathcal T$ equals
  \beq\label{q162}
 \begin{array}{c}
  \displaystyle{
\mathcal T_{N=2}=\left(
\begin {array}{cc} 1&0\\\noalign{\medskip}\tau&1\end {array}
 \right)\,,
\ \ \ \ \
 \mathcal T_{N=3}= \left( \begin {array}{ccc}
1&0&0\\\noalign{}0&1&0
\\\noalign{}0&3\,\tau&1\end {array} \right)\,,
 }
 \ \ \ \ \
 \mathcal T_{N=4}= \left(
\begin {array}{cccc} 1&0&0&0\\\noalign{}0&1&0&0
\\\noalign{}\tau&0&1&0\\\noalign{}6\,{\tau}^{2}&0&6\,
\tau&1\end {array} \right)\,.

 \end{array}
 \eq

 It follows from (\ref{q151})-(\ref{q154}) and (\ref{q159}) that $\tau$-deformation
 of $R$-matrix is given by the following gauge transformation:
  \beq\label{q163}
 \begin{array}{c}
  \displaystyle{
 R^{\hbar}(z|\,\tau)= \mathcal T_1 \mathcal T_2\, R^{\hbar}(z|\,0)\, \mathcal T_1^{-1}\mathcal
 T_2^{-1}
 }
 \end{array}
 \eq
written in terms of (\ref{q1611}). See Appendix A for explicit
answer in ${\rm gl}_3$ case.

\subsection{Rational KZB equations}
It follows from (\ref{q163}) that
  \beq\label{q164}
 \begin{array}{c}
  \displaystyle{
 r_{ab}^\tau(z_a-z_b)= \mathcal T_a \mathcal T_b\, r_{ab}(z_a-z_b)\, \mathcal T_a^{-1}\mathcal
 T_b^{-1}\,,
 }
 \\ \ \\
   \displaystyle{
 m_{ab}^\tau(z_a-z_b)= \mathcal T_a \mathcal T_b\, m_{ab}(z_a-z_b)\, \mathcal T_a^{-1}\mathcal
 T_b^{-1}\,.
 }
 \end{array}
 \eq
 Then the condition (\ref{q111}) is fulfilled as well as (\ref{q1461}) for (\ref{q1371})-(\ref{q1372}).

The Lax pair (\ref{q1041})-(\ref{q1042}) is transformed by not only
the gauge transformation since the residue $S$ also changes. From
(\ref{q1041})-(\ref{q1042}) and (\ref{q164}) we have
  \beq\label{q165}
 \begin{array}{c}
  \displaystyle{
L(z,S,\,\tau)=\mathcal T\, L(z,\mathcal T^{-1}S\mathcal T,\,0)\,
\mathcal T^{-1}\,,
 }
 \end{array}
 \eq
  \beq\label{q166}
 \begin{array}{c}
  \displaystyle{
\mathcal M(z,S,\,\tau)=\mathcal T\, \mathcal M(z,\mathcal
T^{-1}S\mathcal T,\,0)\, \mathcal T^{-1}\,.
 }
 \end{array}
 \eq
Let us summarize the results:
\begin{predl}
The $\tau$-deformed quantum $R$-matrix (\ref{q163}) satisfies the
 Painlev\'e-Calogero pro\-perty (\ref{q1117}).
\end{predl}

\begin{predl}
The $\tau$-deformed quantum $r$ and $m$-matrices (\ref{q164}) define
the KZB equations (\ref{q140}), i.e. the corresponding KZB
connections $\nabla_a$ (\ref{q144}) and   $\nabla_\tau$ (\ref{q145})
are compatible (\ref{q141}), (\ref{q142}).
\end{predl}
The proof is direct. Below we give explicit examples of
$\tau$-deformations in the rational case.

\subsection{Example: ${\rm gl}_2$ case}

Quantum $R$-matrix (satisfying (\ref{q1117})):
 \beq\label{q170}
 \begin{array}{c}
  \displaystyle{R^{\hbar,\tau}(z)=
 \left( \begin{array}{cccc} {\hbar}^{-1}+{z}^{-1}&0&0&0
\\\noalign{\medskip}-\hbar-z&{\hbar}^{-1}&{z}^{-1}&0\\\noalign{\medskip}
-\hbar-z&{z}^{-1}&{\hbar}^{-1}&0\\\noalign{\medskip}-(z+\hbar)(z^2+
z\hbar+\hbar^2+4\tau)&\hbar+z&\hbar+z&{\hbar}^{-1}+{z}^{-1}
\end{array} \right)
  }
 \end{array}
 \eq
Classical $r$-matrix
 \beq\label{q171}
   \displaystyle{
{r}^\tau_{12}(z)=
 \left(\begin{array}{cccc}
\,z^{-1} & 0 & 0 & 0\\ -z & 0 & \,z^{-1} & 0\\ -z & \,z^{-1} & 0 & 0\\
-z^3-4z\tau & z & z & \,z^{-1}
 \end{array}
 \right)
 }
 \eq
and $m$-matrix (the next term of expansion of (\ref{q170}) in
$\hbar$) satisfying (\ref{q111}):
 \beq\label{q172}
   \displaystyle{
{m}^\tau_{12}(z)=
  \left( \begin {array}{cccc} 0&0&0&0\\\noalign{}-1&0&0&0
\\\noalign{}-1&0&0&0\\\noalign{}-2{z}^{2}-4\tau&1&
1&0\end {array} \right)
 }
 \eq
The following additional relation holds:
 \beq\label{q1721}
   \displaystyle{
-\p_z
{r}^\tau_{12}(z)=\frac{P_{12}}{z^2}-\frac{3}{2}\,{m}^\tau_{12}(z)+\frac{1}{2}\,{m}^\tau_{12}(0)\,.
 }
 \eq
Non-autonomous top Lax pair and Hamiltonian:
 \beq\label{q173}
 \begin{array}{c}
  \displaystyle{
L(z,S|\,\tau)=  \frac{1}{z}\left( \begin{array}{cc} S_{11}-z^2S_{12}
& S_{12}
\\ \ \\ 
S_{21}-z^2 (S_{11}-S_{22})-z^4S_{12}-4z^2\tau S_{12} & \
S_{22}+z^2S_{12}
\end{array} \right)
  }
 \end{array}
 \eq
 \beq\label{q174}
 \begin{array}{c}
  \displaystyle{
{\mathcal M}(z,S|\,\tau)=-\mat{ S_{12}}{0}{
S_{11}-S_{22}+2z^2S_{12}+4\tau S_{12}}{-S_{12}}
  }
 \end{array}
 \eq

  \beq\label{q175}
 \begin{array}{c}
  \displaystyle{
H(S,\tau)=-S_{12}(S_{11}-S_{22})-2\tau S_{12}^2\,.
  }
 \end{array}
 \eq
The Gaudin (or Schlesinger) Hamiltonians:
 \beq\label{q176}
 \begin{array}{c}
  \displaystyle{
  h_a=\sum\limits_{c\neq a}^{\ti N} h_{a,c}\,,\ \ \
  h_{a,c}=-\tr_{12}\left(r^\tau_{12}(z_a-z_c) S^a_1  S^c_2\right)=
 }
 \end{array}
 \eq
$$
  \displaystyle{
-\frac{\tr( S^a  S^c)}{z_a-z_c}
  +(z_a-z_c)\Big( S_{12}^a( S_{11}^c- S_{22}^c)+ S_{12}^c( S_{11}^a- S_{22}^a)+4\tau S_{12}^a S_{12}^c\Big)
  +(z_a-z_c)^3\, S_{12}^a S_{12}^c\,,
 }
$$

 \beq\label{q177}
 \begin{array}{c}
  \displaystyle{
h_0=\frac{1}{2}\sum\limits_{b,c\,=1}^{\ti N}\tr\left(S^b\,{ \mathcal
M}(z_b-z_c,S^c)\right)=-\sum\limits_{b,c\,=1}^n
S^b_{12}(S_{11}^c-S_{22}^c) + S_{12}^b S_{12}^c
\left[(z_b-z_c)^2+2\tau\right].
 }
 \end{array}
 \eq
Some similar formulae for ${\rm gl}_3$ case are given in the
Appendix A.




\section{Planck constant as spectral parameter} 
\setcounter{equation}{0}

\subsection{$R$-matrix valued Fay identities}

In this paragraph we show that the quantum $R$-matrices satisfy a
set of relations which are similar to their scalar analogues -- the
functions $\Phi$ (\ref{q1021}). It is convenient to discuss the
elliptic case (\ref{q1480})-(\ref{q1498}) because the trigonometric
and rational versions are obtained by some (nontrivial)
degenerations.

The function $\phi(x,z)$ (\ref{q1480}) (or  (\ref{q1498})) satisfies
the Fay identity:
  \beq\label{q701}
 \begin{array}{c}
  \displaystyle{
 \phi(x,z_{ab})\phi(y,z_{bc})=\phi(x-y,z_{ab})\phi(y,z_{ac})+\phi(y-x,z_{bc})\phi(x,z_{ac})\,,
 }
 \end{array}
 \eq
where $z_{ab}=z_a-z_b$. Let us formulate its noncommutative
analogue.
 \begin{predl}
 The Belavin's $R$-matrix (\ref{q1483}) satisfies the following
 relation:
  \beq\label{q702}
 \begin{array}{c}
  \displaystyle{
 R^\hbar_{ab}
 R^{\hbar'}_{bc}=R^{\hbar'}_{ac}R_{ab}^{\hbar-\hbar'}+R^{\hbar'-\hbar}_{bc}R^\hbar_{ac}\,,
 }
 \end{array}
 \eq
 where $R^\hbar_{ab}=R^\hbar_{ab}(z_a-z_b)$.
 \end{predl}
\noindent\underline{\emph{Proof:}} Denote by $T^a_\al$ the basis
element $T_\al$ (\ref{q1478}) standing on the $a$-th place in the
tensor product $1\otimes...\otimes 1\otimes T_\al\otimes
1\otimes...\otimes 1$. It follows from the definition (\ref{q1483})
and the multiplication rule (\ref{q14791}) that
  \beq\label{q703}
 \begin{array}{c}
  \displaystyle{
 R^\hbar_{ab}
 R^{\hbar'}_{bc}=\sum\limits_{\al,\be} T^a_\al\, T^b_{\be-\al}\,T^c_{-\be}\,
 \kappa_{-\al,\be}\, \vf_\al^\hbar(z_{a}-z_b)\vf_\be^{\hbar'}(z_b-z_c)  \,,
 }
 \end{array}
 \eq
  \beq\label{q704}
 \begin{array}{c}
  \displaystyle{
 R^{\hbar'}_{ac}
 R^{\hbar-\hbar'}_{ab}=\sum\limits_{\al,\be} T^a_\al\, T^b_{\be-\al}\,T^c_{-\be}\,
 \kappa_{\be,\al-\be}\, \vf_\be^{\hbar'}(z_{a}-z_c)\vf_{\al-\be}^{\hbar-\hbar'}(z_a-z_b)  \,,
 }
 \end{array}
 \eq
  \beq\label{q705}
 \begin{array}{c}
  \displaystyle{
 R^{\hbar'-\hbar}_{bc}
 R^{\hbar}_{ac}=\sum\limits_{\al,\be} T^a_\al\, T^b_{\be-\al}\,T^c_{-\be}\,
 \kappa_{\be-\al,\al}\, \vf_{\be-\al}^{\hbar'-\hbar}(z_{b}-z_c)\vf_{\al}^{\hbar}(z_a-z_c)  \,,
 }
 \end{array}
 \eq
Notice that
$\kappa_{-\al,\be}\!=\!\kappa_{\be,\al-\be}\!=\!\kappa_{\be-\al,\al}$
due to (\ref{q14792}). Then the statement (\ref{q702}) follows from
(\ref{q701}), where $x=\hbar+\om_\al$ and $y=\hbar'+\om_\be$.
$\blacksquare$

 \begin{predl}
The quantum Yang-Baxter equation (\ref{q102}) follows from
(\ref{q702}), the property (\ref{q1492}) and unitarity condition
(\ref{q1021}).
 \end{predl}
\noindent\underline{\emph{Proof:}}  Consider (\ref{q702}) for
$a,b,c=1,2,3$ and $\hbar'=\hbar/2$:
 $$
 R_{12}^\hbar
R_{23}^{\hbar/2}=R_{13}^{\hbar/2}R_{12}^{\hbar/2}+R_{23}^{-\hbar/2}R_{13}^\hbar
$$
Replace $\hbar\to2\hbar$ and multiply this relation by
$R_{23}^\hbar$ from the left:
  \beq\label{q7051}
 \begin{array}{c}
  \displaystyle{
 R_{23}^\hbar R_{13}^\hbar R_{12}^\hbar=R_{23}^\hbar
 R_{12}^{2\hbar}R_{23}^\hbar-R_{23}^\hbar
 R_{23}^{-\hbar}R_{13}^{2\hbar}\,.
 }
 \end{array}
 \eq
 Similarly, consider (\ref{q702}) for $a,b,c=1,3,2$ and
 $\hbar'=\hbar/2$, replace $\hbar\to2\hbar$ and multiply the obtained relation by $R_{23}^\hbar$
from the right:
  \beq\label{q7052}
 \begin{array}{c}
  \displaystyle{
 R_{12}^\hbar R_{13}^\hbar R_{23}^\hbar=R_{13}^{2\hbar}
 R_{32}^{\hbar}R_{23}^\hbar-R_{32}^{-\hbar}
 R_{12}^{2\hbar}R_{23}^{\hbar}\,.
 }
 \end{array}
 \eq
 The r.h.s of (\ref{q7051}) equals r.h.s of (\ref{q7052}) due to the
 property (\ref{q1492}) and unitarity condition (\ref{q1021}).$\blacksquare$

Consider the derivative of (\ref{q702}) with respect to $z_b$:
  \beq\label{q706}
 \begin{array}{c}
  \displaystyle{
 R^\hbar_{ab}
 F^{\hbar'}_{bc}-F^\hbar_{ab}
 R^{\hbar'}_{bc}=F^{\hbar'-\hbar}_{bc}R^\hbar_{ac}-R^{\hbar'}_{ac}F_{ab}^{\hbar-\hbar'}\,,
 }
 \end{array}
 \eq
where $F^\hbar_{ab}(z)=\p_z R^\hbar_{ab}(z)$. The function
$F^\hbar_{ab}(z)$ has no singularities at $\hbar=0$. Therefore, we
can put $\hbar=\hbar'$ in (\ref{q706}). This gives
  \beq\label{q707}
 \begin{array}{c}
  \displaystyle{
 R^\hbar_{ab}
 F^{\hbar}_{bc}-F^\hbar_{ab}
 R^{\hbar}_{bc}=F^{0}_{bc}R^\hbar_{ac}-R^{\hbar}_{ac}F_{ab}^{0}\,,
 }
 \end{array}
 \eq
The latter equation is analogue of the following identity
  \beq\label{q708}
 \begin{array}{c}
  \displaystyle{
 \phi(x,z_{ab})f(x,z_{bc})-f(x,z_{ab})\phi(x,z_{bc})=
 \phi(x,z_{ac})(\wp(z_{ab})-\wp(z_{bc}))\,,
 }
 \\ \ \\
  \displaystyle{
 f(x,z_{ab})=\p_{z_a}\phi(x,z_{ab})
 }
 \end{array}
 \eq
underlying Lax equations (integrability) of the Calogero-Moser model
\cite{Calogero,Krich1}.

\subsection{$R$-matrix valued linear problem for Calogero-Moser model}

Consider the eigenvalue problem
  \beq\label{q300}
 \begin{array}{c}
  \displaystyle{
\mathcal L\Psi=\Psi\Lambda
 }
 \end{array}
 \eq
for the following block matrix operator
  \beq\label{q301}
 \begin{array}{c}
  \displaystyle{
\mathcal L=\sum\limits_{a,b=1}^{\ti N} \ti{\mathrm E}_{ab}\otimes
\mathcal L_{ab}\,,
 }
 \end{array}
 \eq
where $\ti{\mathrm E}_{ab}$ is the standard basis of ${\rm gl}_{\ti
N}$ and
  \beq\label{q302}
 \begin{array}{c}
  \displaystyle{
\mathcal L_{ab}=\delta_{ab}p_a\,1_a\otimes
1_b+\nu(1-\delta_{ab})R_{ab}^\hbar\,,\ \ \
R_{ab}^\hbar=R_{ab}^\hbar(z_a-z_b)\,.
 }
 \end{array}
 \eq
 It is worth mentioning that in ${\rm gl}_1$ case ($N=1$) this
 operator is the Krichever's Lax matrix with spectral parameter for
 the
 Calogero-Moser model \cite{Krich1}. The eigenvalue matrix consists
 of vectors $\psi_1,...,\psi_{\ti N}$. In the case of quantum CM model ($p_a\to\p_{z_a}$) equation (\ref{q300})
 should have
 well defined limit $\hbar\to 0$ which gives the KZ equations for $\psi_1=...=\psi_{\ti
 N}=\psi$. Alternatively, one can quantize the model as
 $p_a\to\nabla_a$. At the level of classical mechanics and $N=1$ the
 difference between $\p_{z_a}$ and $\nabla_a$ is given by the canonical
 map $p_a\rightarrow p_a+\nu\sum\limits_{c\neq a}E_1(z_a-z_c)$.

 The spectral parameter in (\ref{q302}) is
 $\hbar$ - the Planck constant. The $M$-operator is defined as
 follows:
  \beq\label{q320}
 \begin{array}{c}
  \displaystyle{
\mathcal M_{ab}=\nu\delta_{ab}
d_a+\nu(1-\delta_{ab})F_{ab}^\hbar+\nu\delta_{ab}\,\mathcal F^0\,,
 }
 \end{array}
 \eq
 where
  \beq\label{q321}
 \begin{array}{c}
  \displaystyle{
F_{ab}^\hbar=\p_{z_a}R_{ab}^\hbar(z_a-z_b)\,,
 }
 \end{array}
 \eq
  \beq\label{q322}
 \begin{array}{c}
  \displaystyle{
d_a=-\sum\limits_{c:\,c\neq a}^{\ti N}F^0_{ac}\,,\ \ \
F^0_{ac}=F^\hbar_{ac}\left.\right|_{\hbar=0}\,,
 }
 \end{array}
 \eq
  \beq\label{q323}
 \begin{array}{c}
  \displaystyle{
\mathcal F^0=\frac{1}{2}\sum\limits_{b,c:\,b\neq c}^{\ti
N}F^0_{bc}=\sum\limits_{b,c:\,b> c}^{\ti N}F^0_{bc}\,.
 }
 \end{array}
 \eq
$M$-operator  (\ref{q320}) is also straightforward generalization of
the one proposed in \cite{Krich1} except the last term $\mathcal
F^0$. The latter is not needed in $N=1$ case because in this case it
is proportional to the identity matrix.

\begin{predl}\label{CM}
The linear problem
  \beq\label{q324}
 \begin{array}{c}
  \displaystyle{
(\p_t+\mathcal M)\Psi=0\,,\ \ \
 \mathcal M=\sum\limits_{a,b=1}^{\ti
N} \ti{\mathrm E}_{ab}\otimes \mathcal M_{ab}
 }
 \end{array}
 \eq
is compatible with (\ref{q300}). The compatibility condition is
equivalent to dynamics of ${\rm gl}_{\ti N}$ Calogero-Moser model.
\end{predl}
\noindent\underline{\emph{Proof:}} The compatibility condition is
the Lax equation $\p_t\mathcal L=[\mathcal L,\mathcal M]$. For
brevity sake let us denote $\mathcal L=p+R$, $\mathcal
M=d+F+\mathcal F^0$.
The commutator equals
  \beq\label{q325}
 \begin{array}{c}
  \displaystyle{
[\mathcal L,\mathcal M]=[p,F]+[R,d]+[R,F]+[R,\mathcal F^0]\,.
 }
 \end{array}
 \eq
The term $[p,F]$ is cancelled by $\p_t R$ (due to $\dot z_a=p_a$).

Consider the off-diagonal block $ac$. It has three inputs from

1. from $[R,F]$:  $\sum\limits_{b\neq a,c}R^\hbar_{ab}
 F^{\hbar}_{bc}-F^\hbar_{ab}
 R^{\hbar}_{bc}\stackrel{(\ref{q707})}{=}\sum\limits_{b\neq a,c}
 F^{0}_{bc}R^\hbar_{ac}-R^{\hbar}_{ac}F_{ab}^{0}$;

2. from $[R,d]$: $-R_{ac}^\hbar\sum\limits_{b\neq
c}F_{cb}^0+\sum\limits_{b\neq a}F_{ab}^0R_{ac}^\hbar$;

3. from $[R,\mathcal F^0]$: $[\mathcal L_{ac},\mathcal F^0]$.

\noindent The sum of the inputs equals zero. We used that
$F_{ab}^0=F_{ba}^0$ (due to $F_{ab}^0=\p_{z_a}r_{ab}(z_a-z_b)$).

On a diagonal block we get equations of motion:
  \beq\label{q326}
 \begin{array}{c}
  \displaystyle{
\dot p_a=\nu^2\sum\limits_{b\neq a}R_{ab}^\hbar
F_{ba}^\hbar-F_{ab}^\hbar
R_{ba}^\hbar\stackrel{(\ref{q305})}{=}N^2\nu^2\sum\limits_{b\neq
a}\wp'(z_a-z_b)\,.
 }
 \end{array}
 \eq
\noindent $\blacksquare$

It is natural to expect that the same receipt works for other root
systems (not only ${\rm gl}_N$) as well, i.e. one can replace the
function $\phi(x,z)$ in the Lax matrix with the corresponding
quantum $R$-matrix.

 Denote the off-diagonal part of (\ref{q302}) by $\mathcal L^0$:
 $
\mathcal L_{ab}^0=(1-\delta_{ab})R_{ab}^\hbar$.
 We conjecture that\footnote{The proof will be given elsewhere.}:
   \beq\label{q304}
 \begin{array}{c}
  \displaystyle{
\tilde\tr\mathcal ((\mathcal
L^0)^{k+1})_{aa}=\!\sum\limits_{b_1,...,b_k=1}^{\ti N}
\! R_{ab_1}^\hbar\,...\,R_{b_ka}^\hbar=1_1\otimes...\otimes1_{\ti N}
\sum\limits_{b_1,...,b_k=1}^{\ti N}
\!\Phi^\hbar(z_{a}-z_{b_1})...\Phi^\hbar(z_{b_k}-z_{a})\,,
 }
 \end{array}
 \eq
where $\ti\tr$ denotes the trace over ${\mathrm gl}_{\ti N}$
component of $\mathcal L$ and the sums do not contain zero arguments
(i.e. $b_1\neq a$, $b_2\neq b_1$, ... ,$b_k\neq a$). Relation
(\ref{q304}) means that traces of $\mathcal L$
(\ref{q301})-(\ref{q302}) provides the Hamiltonians of the ${\rm
gl}_{\ti N}$ Calogero-Moser model (where $z_a$ are coordinates of
particles).

For $k=1$ (\ref{q304}) follows from the unitarity condition:
   \beq\label{q305}
 \begin{array}{c}
  \displaystyle{
\sum\limits_{b}R_{ab}^\hbar\, R_{ba}^\hbar = 1_a\otimes
1_b\sum\limits_{b}\Phi^\hbar(z_a-z_b)\Phi^\hbar(z_b-z_a)=N^2\wp(N\hbar)-N^2\wp(z_a-z_b)\,.
 }
 \end{array}
 \eq
For $k=2$ and $\ti N=3$ we have
   \beq\label{q306}
 \begin{array}{c}
  \displaystyle{
R_{ab}^\hbar\, R^\hbar_{bc}\, R^\hbar_{ca}
\!+\!R^\hbar_{ac}\,R^\hbar_{cb}\,R^\hbar_{ba}=1_a\!\otimes\!
1_b\!\otimes\!
1_c\left(\Phi^\hbar(z_{ab})\Phi^\hbar(z_{bc})\Phi^\hbar(z_{ca})+\Phi^\hbar(z_{ac})\Phi^\hbar(z_{cb})\Phi^\hbar(z_{ba})\right)
 }
 \end{array}
 \eq
($z_{ab}=z_a-z_b$) or, in particular
   \beq\label{q307}
 \begin{array}{c}
  \displaystyle{
R_{12}^\hbar\, R^\hbar_{23}\, R^\hbar_{31}
+R^\hbar_{13}\,R^\hbar_{32}\,R^\hbar_{21}=1\!\otimes\! 1\!\otimes\!
1\left(\Phi^\hbar(z_{12})\Phi^\hbar(z_{23})\Phi^\hbar(z_{31})+\Phi^\hbar(z_{13})\Phi^\hbar(z_{32})\Phi^\hbar(z_{21})\right)
 }
 \end{array}
 \eq
 The function in the r.h.s. of (\ref{q307})
equals
   \beq\label{q308}
 \begin{array}{c}
  \displaystyle{
\Phi^\hbar(z_{12})\Phi^\hbar(z_{23})\Phi^\hbar(z_{31})+\Phi^\hbar(z_{13})\Phi^\hbar(z_{32})\Phi^\hbar(z_{21})=
  \left\{
  \begin{array}{l}
  -N^3\wp'(\hbar)\ \hbox{in}\ \hbox{elliptic}\ \hbox{case}\,,
  \\ \ \\
  2/\hbar^3\ \hbox{in}\ \hbox{rational}\ \hbox{case}\,.
  \end{array}
  \right.
 }
 \end{array}
 \eq

\subsection{Half of the classical Yang-Baxter equation}


Consider the unitarity condition
$R_{ab}^\hbar\,R_{ba}^\hbar=\Phi^\hbar(z_{ab})\Phi^\hbar(z_{ba})$.
Its expansion in the $\hbar^0$ order gives
 \beq\label{q404}
 \begin{array}{c}
  \displaystyle{
r_{ab}^2-2m_{ab}=1_a\otimes 1_b\, N^2\wp(z_a-z_b)\,.
 }
 \end{array}
 \eq
 Here
$r_{ab}=r^\tau_{ab}(z_a-z_b)$, $m_{ab}=m^\tau_{ab}(z_a-z_b)$. Next,
consider (\ref{q306})-(\ref{q308}). In the $\hbar^1$ order it
provides the following relation between $r$ and $m$ matrices:
 \beq\label{q405}
 \begin{array}{c}
  \displaystyle{
[r_{ab},r_{bc}]_+ +[r_{bc},r_{ca}]_+ +[r_{ab},r_{ca}]_+
+2(m_{ab}+m_{bc}+m_{ac})=0\,,
 }
 \end{array}
 \eq
where $[*,*]_+$ is the anticommutator $[A,B]_+:=AB+BA$. Using the
classical Yang-Baxter equation
 \beq\label{q406}
 \begin{array}{c}
  \displaystyle{
[r_{ab},r_{ac}]+[r_{ac},r_{bc}]+[r_{ab},r_{bc}]=0
 }
 \end{array}
 \eq
 we can combine (\ref{q405}) and  (\ref{q406}) into two "halves" of
 the classical
Yang-Baxter equation:
 \beq\label{q407}
 \begin{array}{c}
  \displaystyle{
r_{ab}\,r_{ac}-r_{bc}\,r_{ab}+r_{ac}\,r_{bc}=m_{ab}+m_{bc}+m_{ac}
 }
 \end{array}
 \eq
and
 \beq\label{q408}
 \begin{array}{c}
  \displaystyle{
r_{ac}\,r_{ab}-r_{ab}\,r_{bc}+r_{bc}\,r_{ac}=m_{ab}+m_{bc}+m_{ac}\,.
 }
 \end{array}
 \eq
The difference of (\ref{q407}) and  (\ref{q408}) gives (\ref{q406})
while the sum leads to (\ref{q405}).

In the light of (\ref{q404}) the expansion
$R^\hbar(z)=\hbar^{-1}+r(z)+\hbar\, m(z)$ is similar to the
expansion (\ref{q150}). Indeed, using (\ref{q404}) we have
  \beq\label{q1499}
 \begin{array}{c}
  \displaystyle{
R_{ab}^\hbar(z)=\frac{1}{\hbar}\,1_a\otimes
1_b+r_{ab}+\hbar\,m_{ab}+...=\frac{1}{\hbar}\,1_a\otimes
1_b+r_{ab}+\frac{\hbar}{2}\left(r_{ab}^2-N^2\wp(z_{ab})\right)+...\,.
 }
 \end{array}
 \eq
In the same time (\ref{q405}) can be re-written as
 \beq\label{q409}
 \begin{array}{c}
  \displaystyle{
(r_{ab}+r_{bc}+r_{ca})^2=1_a\otimes 1_b\otimes 1_c\,
N^2(\wp(z_a-z_b)+\wp(z_b-z_c)+\wp(z_c-z_a))
 }
 \end{array}
 \eq
using (\ref{q404}). It is an analogue of the elliptic functions
identity
 \beq\label{q410}
 \begin{array}{c}
  \displaystyle{
(E_1(z_a-z_b)+E_1(z_b-z_c)+E_1(z_c-z_a))^2=\wp(z_a-z_b)+\wp(z_b-z_c)+\wp(z_c-z_a)\,.
 }
 \end{array}
 \eq
%
%


\subsection{Identities for KZB equations}

 It follows from
(\ref{q404}) that
  \beq\label{q403}
 \begin{array}{c}
  \displaystyle{
[r_{ab},m_{ab}]=0\,.
 }
 \end{array}
 \eq
 This is equation (\ref{q147}) written in the fundamental
 representation (in this case $m_{aa}$ are some scalar operators).
 Equation (\ref{q148}) keeps its form in the fundamental
 representation.
Let us prove it.
\begin{predl}
 The following identities holds true:
 \beq\label{q411}
 \begin{array}{c}
  \displaystyle{
 [r_{ab},m_{ac}+m_{bc}]+[r_{ac},m_{ab}+m_{bc}]=0\,,
 }
 \end{array}
 \eq
  \beq\label{q412}
 \begin{array}{c}
  \displaystyle{
 [r_{bc},m_{ab}-m_{ac}]+r_{ab}r_{bc}r_{ac}-r_{ac}r_{bc}r_{ab}=0\,.
 }
 \end{array}
 \eq
 The first one underlies the compatibility of KZB equations. See
 (\ref{q148}).
\end{predl}

\noindent\underline{\emph{Proof:}} Consider the Yang-Baxter equation
$R_{ca}^\hbar R_{cb}^\hbar R_{ab}^\hbar=R_{ab}^\hbar R_{cb}^\hbar
R_{ca}^\hbar$ in the $\hbar^0$ order. It is given by the sum of
(\ref{q411}) and (\ref{q412}). Consider also (\ref{q306}) in the
$\hbar^0$ order. It is given by the difference of (\ref{q411}) and
(\ref{q412}). $\blacksquare$

The identities (\ref{q404})-(\ref{q405}) allow also to get the
following Matsuo-Cherednik's like \cite{Matsuo,Cherednik2}
statement:


 \begin{predl}
Consider the ${\rm gl}_N$ KZB equations for $\ti N$ punctures:
 \beq\label{q200}
 \begin{array}{c}
  \displaystyle{
\nabla_i\psi=0\,, \ \ \nabla_i=\p_i+\nu\sum\limits_{j:j\neq i}
r_{ij}^\tau(z_i-z_j)\,,
 }
 \end{array}
 \eq
for $i=1,...,\ti N$ and\footnote{The summation of indices runs over
$1...{\ti N}$. Here and elsewhere we shall omit the  limits of
summation when it can be done without ambiguity.}
  \beq\label{q201}
 \begin{array}{c}
  \displaystyle{
\nabla_\tau\psi=0\,, \ \
\nabla_i=\p_\tau+\frac{\nu}{2}\sum\limits_{j\neq k}
m_{jk}^\tau(z_j-z_k)\,,
 }
 \end{array}
 \eq
where $r^\tau_{ij}$ and $m^\tau_{ij}$ are the coefficients of the
expansion (\ref{q104}) and $\nu$ is a free constant. Then the
conformal block satisfies the following equation:
  \beq\label{q220}
 \begin{array}{c}
  \displaystyle{
\left(\ti N\nu\p_\tau +\frac{1}{2}\,\Delta\right)\psi=\left(
-\nu\sum\limits_{i<j}\p_i r^\tau_{ij}-\frac{1}{2}\ti
N\nu^2\,\sum\limits_{j} m^\tau_{jj}+\nu^2 N^2
\sum\limits_{i<j}1_i\otimes 1_j\,\wp(z_i-z_j)\right)\psi
 }
 \end{array}
 \eq
where $\Delta=\sum\limits_i \p_i^2$ and $m_{jj}^\tau=m_{jj}^\tau(0)$
are scalar operators depending on $\tau$.
 \end{predl}

\noindent\underline{\emph{Proof:}}
Let us omit the dependence on $\tau$, i.e. $r^\tau_{ij}:=r_{ij}$.
 \beq\label{q205}
 \begin{array}{c}
  \displaystyle{
\p_i^2\psi=\left(-\nu\sum\limits_{j:j\neq i}\p_i
r_{ij}+\nu^2\left(\sum\limits_{j:j\neq i}
r_{ij}\right)^2\,\right)\psi\,.
 }
 \end{array}
 \eq
Summing up equations (\ref{q205}) for $i=1...\ti N$ we get
 \beq\label{q206}
 \begin{array}{c}
  \displaystyle{
\frac{1}{2}\,\Delta\,\psi=\left( -\nu\sum\limits_{i<j}\p_i
r_{ij}+\nu^2\sum\limits_{i<j}r_{ij}^2+\frac{1}{2}\,\nu^2\,\sum\limits_{k}\sum\limits_{i<j}\,
[r_{ki},r_{kj}]_+\right)\psi
 }
 \end{array}
 \eq
Let us transform the last sum using identity (\ref{q405}):
 \beq\label{q207}
 \begin{array}{c}
  \displaystyle{
\frac{1}{2}\,\sum\limits_{k}\sum\limits_{i<j}\,
[r_{ki},r_{kj}]_+=-\frac{1}{2}\,\sum\limits_{k<i<j}\,
[r_{ki},r_{ij}]_++[r_{ij},r_{jk}]_++[r_{jk},r_{ki}]_+
 }
 \\ \ \\
  \displaystyle{
\stackrel{(\ref{q405})}{=} \sum\limits_{k<i<j}\,\left(
m_{ki}+m_{kj}+m_{ij} \right)=(\ti N-2)\,\sum\limits_{i<j} m_{ij}
 }
 \end{array}
 \eq
Plugging it into the r.h.s. of (\ref{q206}) and using  (\ref{q404})
we obtain:
 \beq\label{q208}
 \begin{array}{c}
  \displaystyle{
\frac{1}{2}\,\Delta\,\psi=\left( -\nu\sum\limits_{i<j}\p_i
r_{ij}+\nu^2\sum\limits_{i<j}r_{ij}^2+(\ti
N-2)\,\nu^2\,\sum\limits_{i<j} m_{ij}\right)\psi
 }
\\ \ \\
  \displaystyle{
\stackrel{(\ref{q404})}{=} \left( -\nu\sum\limits_{i<j}\p_i
r_{ij}+\nu^2 N^2 \sum\limits_{i<j}1_i\otimes 1_j\,\wp(z_i-z_j)+\ti
N\nu^2\,\sum\limits_{i<j} m_{ij}\right)\psi\,.
 }
 \end{array}
 \eq
The Proposition result (\ref{q220}) follows from (\ref{q208}) and
(\ref{q201}). $\blacksquare$

\subsection{Painlev\'e equations}

The block matrix Lax pair (\ref{q302}), (\ref{q320}) can be also
used for description of the Painlev\'e equations likewise it was
done in \cite{LO} in $N=1$ case, i.e. the result of Proposition
\ref{CM} is naturally generalized to the following one:
 \begin{predl}
 Consider the linear problem
\beq\label{q230}
 \left\{\begin{array}{c}
  \displaystyle{
(\p_\hbar+\mathcal L)\Psi=0\,,
 }
 \\ \ \\
\displaystyle{
 (\p_\tau+\mathcal M)\Psi=0\,,
 }
 \end{array}\right.
 \eq
where $\mathcal L$ and $\mathcal M$ are defined by (\ref{q302}),
(\ref{q320}). The compatibility condition
 \beq\label{q231}
 \begin{array}{c}
  \displaystyle{
\p_\tau \mathcal L-\p_\hbar \mathcal M=[\mathcal L,\mathcal M]
 }
 \end{array}
 \eq
is equivalent to ${\rm gl}_{\ti N}$ Painlev\'e equations
 \beq\label{q235}
 \begin{array}{c}
  \displaystyle{
\p_\tau^2 z_a=N^2\nu^2\sum\limits_{b\neq a}\wp'(z_a-z_b|\tau)\,.
 }
 \end{array}
 \eq
 \end{predl}
The proof repeats the one for the Proposition \ref{CM}. Additionally
one should use the property (\ref{q1117}) of the Painlev\'e-Calogero
correspondence.

\section{Appendix A: ${\rm gl}_3$ (rational) case}
\def\theequation{A.\arabic{equation}}
\setcounter{equation}{0}

Undeformed ${\rm gl}_3$ quantum $R$-matrix:
 \beq\label{q90}
 \begin{array}{c}
  \displaystyle{
  R^\hbar(z)=
  }
 \end{array}
 \eq
{\footnotesize{
 $$
 \left( \begin{array}{ccc} {\hbar}^{-1}+{z}^{-1}&0&0
\\\noalign{\medskip}1&{\hbar}^{-1}&0\\\noalign{\medskip}2\,{\hbar}^{2}+3
\,z\hbar+2\,{z}^{2}&-3\,\hbar-3\,z&{\hbar}^{-1}\\\noalign{\medskip}-1&{z}
^{-1}&0\\\noalign{\medskip}2\,\hbar+2\,z&0&0\\\noalign{\medskip}2\,{z}^
{3}+3\,z{\hbar}^{2}+2\,{\hbar}^{3}+3\,{z}^{2}\hbar&-3\,{\hbar}^{2}-3\,z
\hbar-{z}^{2}&1\\\noalign{\medskip}-2\,{\hbar}^{2}-3\,z\hbar-2\,{z}^{2}&-
3\,\hbar-3\,z&{z}^{-1}\\\noalign{\medskip}2\,{z}^{3}+3\,z{\hbar}^{2}+2\,
{\hbar}^{3}+3\,{z}^{2}\hbar&3\,{z}^{2}+3\,z\hbar+{\hbar}^{2}&-1
\\\noalign{\medskip}2\,{\hbar}^{5}+3\,{z}^{4}\hbar+3\,{z}^{2}{\hbar}^{3}+
2\,{z}^{5}+3\,z{\hbar}^{4}+3\,{z}^{3}{\hbar}^{2}&3\,{z}^{4}-3\,{\hbar}^{4
}-3\,z{\hbar}^{3}+3\,{z}^{3}\hbar&-{z}^{2}+{\hbar}^{2}\end{array}
 \right.
 $$

 $$
 \left. \begin{array}{cccccc} 0&0&0&0&0&0\\\noalign{\medskip}{z}^{-1}
&0&0&0&0&0\\\noalign{\medskip}-3\,\hbar-3\,z&3&0&{z}^{-1}&0&0
\\\noalign{\medskip}{\hbar}^{-1}&0&0&0&0&0\\\noalign{\medskip}0&{\hbar}^
{-1}+{z}^{-1}&0&0&0&0\\\noalign{\medskip}-3\,z\hbar-3\,{z}^{2}-{\hbar}^{
2}&0&{\hbar}^{-1}&1&{z}^{-1}&0\\\noalign{\medskip}-3\,\hbar-3\,z&-3&0&{
\hbar}^{-1}&0&0\\\noalign{\medskip}{z}^{2}+3\,{\hbar}^{2}+3\,z\hbar&0&{z}
^{-1}&-1&{\hbar}^{-1}&0\\\noalign{\medskip}3\,z{\hbar}^{3}+3\,{\hbar}^{4}
-3\,{z}^{3}\hbar-3\,{z}^{4}&-6\,{\hbar}^{3}-6\,{z}^{3}-9\,z{\hbar}^{2}-9
\,{z}^{2}\hbar&3\,z+3\,\hbar&-{\hbar}^{2}+{z}^{2}&3\,z+3\,\hbar&{\hbar}^{-1
}+{z}^{-1}\end {array}\right)
 $$
}}

\vskip2mm

\noindent The $\tau$-deformation generated by (\ref{q163}) with
$\mathcal
 T_{N=3}$ from (\ref{q162}) yields
 \beq\label{q91}
 \begin{array}{c}
  \displaystyle{
  R^{\hbar}(z|\,\tau)=R^{\hbar}(z|\,0)+\delta R^{\hbar,\tau}(z)\,,
  }
  \\ \ \\
  \displaystyle{
  \delta R^{\hbar,\tau}(z)=
  }
 \end{array}
 \eq
{\footnotesize{
 $$
= 3\tau\times\left( \begin {array}{ccccccccc} 0&0&0&0&0&0&0&0&0
\\\noalign{\medskip}0&0&0&0&0&0&0&0&0\\\noalign{\medskip}1&0&0&0&0&0&0
&0&0\\\noalign{\medskip}0&0&0&0&0&0&0&0&0\\\noalign{\medskip}0&0&0&0&0
&0&0&0&0\\\noalign{\medskip}2\,\hbar+2\,z&-1&0&-1&0&0&0&0&0
\\\noalign{\medskip}-1&0&0&0&0&0&0&0&0\\\noalign{\medskip}2\,\hbar+2\,z
&1&0&1&0&0&0&0&0\\\noalign{\medskip}2(z+\hbar)(2\,z^2+z\,\hbar+2\,\hbar^2+3\,\tau)&3\,{z}^{2}-3\,{\hbar}^{
2}&0&-3\,{z}^{2}+3\,{\hbar}^{2}&-6\,z-6\,\hbar&0&0&0&0\end {array}
 \right)
 $$
}}

\noindent Classical $tau$-deformed $r$ and $m$-matrix:

 \beq\label{q92}
 \begin{array}{c}
  \displaystyle{
  r^\tau(z)=
  }
 \end{array}
 \eq
{\footnotesize{
 $$
 \left( \begin {array}{ccccccccc} {z}^{-1}&0&0&0&0&0&0&0&0
\\\noalign{\medskip}1&0&0&{z}^{-1}&0&0&0&0&0\\\noalign{\medskip}3\,
\tau+2\,{z}^{2}&-3\,z&0&-3\,z&3&0&{z}^{-1}&0&0\\\noalign{\medskip}-1&{
z}^{-1}&0&0&0&0&0&0&0\\\noalign{\medskip}2\,z&0&0&0&{z}^{-1}&0&0&0&0
\\\noalign{\medskip}2\,{z}^{3}+6\,\tau\,z&-{z}^{2}-3\,\tau&1&-3\,{z}^{
2}-3\,\tau&0&0&1&{z}^{-1}&0\\\noalign{\medskip}-3\,\tau-2\,{z}^{2}&-3
\,z&{z}^{-1}&-3\,z&-3&0&0&0&0\\\noalign{\medskip}2\,{z}^{3}+6\,\tau\,z
&3\,{z}^{2}+3\,\tau&-1&{z}^{2}+3\,\tau&0&{z}^{-1}&-1&0&0
\\\noalign{\medskip}18\,{\tau}^{2}z+12\,\tau\,{z}^{3}+2\,{z}^{5}&9\,
\tau\,{z}^{2}+3\,{z}^{4}&-{z}^{2}&-9\,\tau\,{z}^{2}-3\,{z}^{4}&-6\,{z}
^{3}-18\,\tau\,z&3\,z&{z}^{2}&3\,z&{z}^{-1}\end {array} \right)
 $$
}}
%
%
{\footnotesize{
 $$
  m^\tau(z)=\left( \begin {array}{ccccccccc} 0&0&0&0&0&0&0&0&0
\\\noalign{\medskip}0&0&0&0&0&0&0&0&0\\\noalign{\medskip}3\,z&-3&0&-3&0
&0&0&0&0\\\noalign{\medskip}0&0&0&0&0&0&0&0&0\\\noalign{\medskip}2&0&0
&0&0&0&0&0&0\\\noalign{\medskip}3\,{z}^{2}+6\,\tau&-3\,z&0&-3\,z&0&0&0
&0&0\\\noalign{\medskip}-3\,z&-3&0&-3&0&0&0&0&0\\\noalign{\medskip}3\,
{z}^{2}+6\,\tau&3\,z&0&3\,z&0&0&0&0&0\\\noalign{\medskip}18\,{\tau}^{2
}+18\,\tau\,{z}^{2}+3\,{z}^{4}&3\,{z}^{3}&0&-3\,{z}^{3}&-9\,{z}^{2}-18
\,\tau&3&0&3&0\end {array} \right)
 $$
}}

\noindent The Lax pair for  $\tau$-deformed (autonomous or
non-autonomous) rational top can be found from
(\ref{q1041})-(\ref{q1042}). It describes dynamics generated by the
following Hamiltonian:
 \beq\label{q94}
 \begin{array}{c}
  \displaystyle{
  H=S_{12}^2-3S_{11}S_{23}+3S_{33}S_{23}-3S_{13}S_{21}+6\tau
  S_{12}S_{13}-9\tau S_{23}^2+9\tau^2 S_{13}^2\,.
  }
 \end{array}
 \eq

\section{Appendix B: Belavin's $R$-matrix}
\def\theequation{B.\arabic{equation}}
\setcounter{equation}{0}

Consider the following basis in ${\rm gl}_N$ (some details can be
found in \cite{LOZ5}):
  \beq\label{q1478}
 \begin{array}{c}
  \displaystyle{
 T_a=T_{a_1 a_2}=\exp\left(\frac{\pi\imath}{N}\,a_1
 a_2\right)Q^{a_1}\Lambda^{a_2}\,,
 }
 \end{array}
 \eq
where $a_1\,,a_2\in {\mathbb Z}_N$ and
  \beq\label{q1479}
 \begin{array}{c}
  \displaystyle{
Q_{kl}=\delta_{kl}\exp(\frac{2\pi
 i}{N}k)\,,\ \ \ \Lambda_{kl}=\delta_{k-l+1=0\,{\hbox{\tiny{mod}}}
 N}\,,\ \ k,l=1,...,N\,.
 }
 \end{array}
 \eq
The multiplication is defied by the following relation:
  \beq\label{q14791}
 \begin{array}{c}
  \displaystyle{
T_{a_1 a_2}T_{b_1 b_2}=\kappa_{a,b}\, T_{a_1+b_1,a_2+b_2}\,,
 }
 \end{array}
 \eq
where
  \beq\label{q14792}
 \begin{array}{c}
  \displaystyle{
\kappa_{a,b}=\exp\left(\frac{\pi \imath}{N}(b_1
a_2-b_2a_1)\right)\,.
 }
 \end{array}
 \eq
For the odd Riemann theta function $\vth(z)=\vth(z|\tau)$
    \beq\label{q1480}
 \begin{array}{c}
  \displaystyle{
\phi(z,u)=\frac {\vth'(0)\vth(u+z)} {\vth(z)\vth(u)}\,,
 }
 \end{array}
 \eq
     \beq\label{q1481}
 \begin{array}{c}
  \displaystyle{
\vf_a(z)=\exp(2\pi\imath z\p_\tau\om_a)\phi(z,\om_a)\,,\ \
\om_a=\frac{a_1+a_2\tau}{N}\,,
 }
 \end{array}
 \eq
     \beq\label{q1482}
 \begin{array}{c}
  \displaystyle{
\vf_a^\hbar(z)=\exp(2\pi\imath z\p_\tau\om_a)\phi(z,\om_a+\hbar)\,.
 }
 \end{array}
 \eq
The Belavin's $R$-matrix \cite{Belavin} can be defined as
   \beq\label{q1483}
 \begin{array}{c}
  \displaystyle{
R^\hbar_{12}(z)=\sum\limits_{\al\in {\mathbb Z}_N\times {\mathbb
Z}_N}\vf_\al^\hbar(z)\,T_\al\otimes T_{-\al}\,.
 }
 \end{array}
 \eq
The local behavior of $\phi(\hbar,z)$ (\ref{q1480}) near $\hbar=0$
is give by
  \beq\label{q150}
 \begin{array}{c}
  \displaystyle{
\phi(\hbar,z)=\frac{1}{\hbar}+E_1(z)+\frac{\hbar}{2}\left(E_1^2(z)-\wp(z)\right)+...\,,
 }
 \end{array}
 \eq
where
   \beq\label{q1484}
 \begin{array}{c}
  \displaystyle{
E_1(z)=\p_z\log\vth(z)
 }
 \end{array}
 \eq
and $\wp(z)$ is the Weierstrass  $\wp$-function. Therefore,
expansion (\ref{q104}) of (\ref{q1483}) gives
   \beq\label{q1495}
 \begin{array}{c}
  \displaystyle{
r_{12}(z)=E_1(z)\,1\otimes 1 +\sum\limits_{\al\neq
0}\vf_\al(z)\,T_\al\otimes T_{-\al}\,,
 }
 \end{array}
 \eq
    \beq\label{q1496}
 \begin{array}{c}
  \displaystyle{
m_{12}(z)=\frac{E_1^2(z)-\wp(z)}{2}\,1\otimes 1
+\sum\limits_{\al\neq 0}f_\al(z)\,T_\al\otimes T_{-\al}\,,
 }
 \end{array}
 \eq
where
    \beq\label{q1497}
 \begin{array}{c}
  \displaystyle{
f_a(z)=\exp(2\pi\imath
z\p_\tau\om_a)\p_u\phi(z,u)\left.\right|_{u=\om_\al}\,.
 }
 \end{array}
 \eq
The function $\Phi$ entering the unitarity condition (\ref{q1021})
equals
    \beq\label{q1498}
 \begin{array}{c}
  \displaystyle{
\Phi^\hbar(z)=N\phi(N\hbar,z)\,.
 }
 \end{array}
 \eq
Notice that the residue of the $R$-matrix (\ref{q1483}) at $z=0$
equals $NP_{12}$, where $P_{12}=N^{-1}\sum\limits_a T_a\otimes
T_{-a}$ is the permutation operator.

It follows from the heat equation for function (\ref{q1482})
    \beq\label{q14991}
 \begin{array}{c}
  \displaystyle{
\p_\tau\vf_a^\hbar(z)=\p_z\p_\hbar \vf_a^\hbar(z)
 }
 \end{array}
 \eq
 that the $R$-matrix (\ref{q1483}) satisfies the property
 (\ref{q1117}):
    \beq\label{q14992}
 \begin{array}{c}
  \displaystyle{
\p_\tau R_{ab}^\hbar=\p_z\p_\hbar R_{ab}^\hbar\,.
 }
 \end{array}
 \eq


     \renewcommand{\refname}{{\normalsize{References}}}


 \begin{small}

 \end{small}

\end{document}